\documentclass[PAPER, atlasdraft=false, UKenglish, cernpreprint, texmf, orcidlogo]{atlasdoc}
\usepackage{atlaspackage}
\usepackage{atlasbiblatex}

\usepackage[BSM=true, hepparticle=true, hepprocess=true, snippets=true]{atlasphysics}
\usepackage[italic]{hepnames}

\usepackage{tikz}
\usetikzlibrary{math}
\usepackage{tikz-feynman,contour}
\usepackage{pgf-pie}
\usepackage{multirow}
\usepackage{makecell}
\usepackage[export]{adjustbox}
\graphicspath{{logos/}{figures/}}

\usepackage{ANA-HDBS-2020-11-PAPER-defs}
\usepackage{MC_snippets-defs} %

\AtlasTitle{Search for a light charged Higgs boson in $t \to \Hpm b$ decays, with $\Hpm \to cs$, in \pp collisions at $\sqrt{s}=\SI{13}{\TeV}$ with the ATLAS detector}

\AtlasAbstract{%
A search for a light charged Higgs boson produced in decays of the top quark, $t \to \Hpm b$ with $\Hpm \to cs$, is presented.
This search targets the production of top-quark pairs $\ttbar \to Wb \Hpm b$, with $W \to \ell\nu$ ($\ell = e, \mu$),
resulting in a lepton-plus-jets final state characterised by an isolated electron or muon and at least four jets.
The search exploits $b$-quark and $c$-quark identification techniques as well as multivariate methods to suppress the dominant \ttbar background.
The data analysed correspond to $\SI{140}{\ifb}$ of \pp collisions at $\sqrt{s} = \SI{13}{\TeV}$ recorded with the ATLAS detector at the LHC between 2015 and 2018.
Observed (expected) \SI{95}{\%} confidence-level upper limits on the branching fraction $\mathscr{B}(t\to \Hpm b)$, assuming $\mathscr{B}(t\to Wb) + \mathscr{B}(t \to \Hpm (\to cs)b)=1.0$,
are set between \SI{0.066}{\%} (\SI{0.077}{\%}) and \SI{3.6}{\%} (\SI{2.3}{\%}) for a charged Higgs boson with a mass between \SI{60}{\GeV} and \SI{168}{\GeV}.
}

\AtlasRefCode{HDBS-2020-11}

\PreprintIdNumber{CERN-EP-2024-185}

\AtlasJournal{EPJC}
\AtlasJournalRef{Eur. Phys. J. C 85 (2025) 153}
\AtlasDOI{10.1140/epjc/s10052-024-13715-4}
\hypersetup{pdftitle={ATLAS document},pdfauthor={The ATLAS Collaboration}}

\begin{document}

\maketitle

\tableofcontents

%
%

%
%

%
%
%

\section{Introduction}
\label{Sec:Introduction}

The discovery of the Higgs boson at the Large Hadron Collider (LHC) in 2012 was a great achievement of the ATLAS and CMS collaborations~\cite{HIGG-2012-27,CMS-HIG-12-028},
and has led to numerous measurements to determine its properties~\cite{HIGG-2021-23,CMS-HIG-22-001}.
One of the main goals of these studies is to establish if the discovered Higgs boson is the single fundamental scalar particle of the
Standard Model (SM) or rather the first observed particle of an extended scalar sector.

Extensions to the scalar sector are motivated by the solutions they provide to several open questions in particle physics.
An extended scalar sector can modify the electroweak phase transition and facilitate baryogenesis~\cite{Riotto:1999yt},
enhance vacuum stability, provide a dark-matter candidate~\cite{Akeroyd:2016ymd,Arcadi:2019lka}
or yield a solution to the strong CP problem~\cite{Peccei:1977hh}.
Many physics models beyond the SM (BSM) require an extended scalar sector.
For example, in the minimal supersymmetric extension of the SM the existence of two Higgs doublets is required~\cite{Martin:1997ns}.
In models with a Type-II seesaw mechanism, Higgs triplets~\cite{PhysRevD.22.2860,PhysRevD.22.2227,LAZARIDES1981287,CHANOWITZ1985105,PhysRevD.42.1673} are required.

Two-Higgs-doublet models (2HDMs)~\cite{Branco:2011iw,Gunion:2002zf} are popular and simple extensions of the scalar sector
and predict the existence of two charged Higgs bosons, \Hplus and \Hminus, and two neutral Higgs bosons in addition to the
discovered neutral one.
The various 2HDMs are categorised into types defined by the Yukawa couplings of the fermions to the Higgs doublets.
The production mechanisms and decay modes of charged Higgs bosons depend on the Yukawa couplings and other
model parameters, especially the ratio of the two Higgs-doublet vacuum expectation values ($\tan{\beta}$) and the
charged Higgs boson's mass ($m_{\Hpm}$).
Many phenomenology studies advocate searching for a light charged Higgs bosons (below the top-quark mass) in the decays
to a charm quark and a strange quark, $\Hpm \rightarrow cs$,\footnote{Unless explicitly stated otherwise, charge conjugation is implied in this paper; the notation $cs$ is used in place of $c\bar{s}/\bar{c}s$.}
to a charm quark and a bottom quark, $\Hpm \rightarrow cb$, and to a $\tau$-lepton and a $\tau$-neutrino, $\Hpm \rightarrow \tau \nu_\tau$~\cite{Branco:2011iw,Akeroyd:2016ymd}.
The branching fraction for $\Hpm \rightarrow cb$ is typically smaller than for $\Hpm \rightarrow cs$, due to the different values of the CKM matrix elements, $V_{cs} \gg V_{cb}$.
In the Type-I 2HDM, only the $\tau \nu_\tau$ and $cs$ decay modes are relevant. In Type-II and Type-X (or \enquote{lepton-specific}) models the $cs$ channel is dominant
for values of $\tan{\beta}<1$. In Type-Y (or \enquote{flipped}) models the $cs$ and $cb$ channels are important for $\tan{\beta}>5$~\cite{Akeroyd:2016ymd}.
For a 2HDM model where one doublet couples mainly to the third generation, while the other doublet couples mainly to the first and second generations,
the $cs$ and $cb$ channels are dominant~\cite{Altmannshofer:2016zrn}. This is also the case for leptophobic multiple-Higgs-doublet models~\cite{Akeroyd:2016ymd}.

Searches for $\Hpm \rightarrow cs$ in top-quark decays have been performed by the ATLAS and CMS collaborations, based on $\SI{4.7}{\ifb}$ and $\SI{35.9}{\ifb}$
of proton--proton (\pp) collision data collected at centre-of-mass energies of $\sqrt{s} = \SI{7}{\TeV}$ and $\sqrt{s} = \SI{13}{\TeV}$, respectively.
The search performed by the ATLAS Collaboration obtained \SI{95}{\%} confidence level (CL) observed (expected) upper limits on the branching fraction
$\mathscr{B} (t \rightarrow \Hpm b)$ ranging from \SI{1.2}{\%} (\SI{1.5}{\%}) to \SI{5.1}{\%} (\SI{8}{\%}), assuming $\mathscr{B} (\Hpm \rightarrow cs) = 1.0$,
for $m_{\Hpm}$ between \SI{90}{\GeV} and \SI{150}{\GeV}~\cite{HIGG-2012-10}.
The search performed by the CMS Collaboration obtained \SI{95}{\%} CL observed (expected) upper limits on $\mathscr{B} (t \rightarrow \Hpm b)$
ranging from \SI{0.25}{\%} (\SI{0.29}{\%}) to \SI{1.68}{\%} (\SI{2.39}{\%}), assuming $\mathscr{B} (\Hpm \rightarrow cs) = 1.0$,
for $m_{\Hpm}$ between \SI{80}{\GeV} and \SI{160}{\GeV}~\cite{CMS-HIG-18-021}.
Related searches for $\Hpm \rightarrow cb$ in top-quark decays were performed by the ATLAS and CMS collaborations using $\SI{139}{\ifb}$ and $\SI{19.7}{\ifb}$ of \pp
collision data collected at $\sqrt{s} = \SI{13}{\TeV}$~\cite{HDBS-2019-24} and \SI{8}{\TeV}~\cite{CMS-HIG-16-030}, respectively. The most stringent observed limits on
$\mathscr{B}(t \rightarrow \Hpm b) \times \mathscr{B}(\Hpm \rightarrow cb)$ are set by ATLAS and range from \SI{0.15}{\%} to \SI{0.42}{\%} for
$m_{\Hpm}$ between \SI{60}{\GeV} and \SI{160}{\GeV}. In that analysis a moderate excess of signal events is observed in the vicinity of \SI{130}{\GeV}, with a global significance of $2.5\sigma$.
Searches for $H^\pm \rightarrow \tau \nu$ in $\SI{36.1}{\ifb}$ of \pp collision data recorded with the ATLAS and CMS detectors at $\sqrt{s} = \SI{13}{\TeV}$ were carried
out over a wide mass range, from \SI{90}{\GeV} to \SI{2000}{\GeV}~\cite{HIGG-2016-11} or from \SI{80}{\GeV} to \SI{3000}{\GeV}~\cite{CMS-HIG-18-014} respectively,
covering the masses of light and heavy charged Higgs bosons. The ATLAS (CMS) Collaboration set observed upper limits on the branching fraction $\mathscr{B}(t \rightarrow \Hpm b) \times
\mathscr{B}(\Hpm \rightarrow \tau\nu)$ ranging from \SI{0.25}{\%} to \SI{0.031}{\%} (\SI{0.36}{\%} to \SI{0.079}{\%}) in the mass range between \SI{90}{\GeV} (\SI{80}{\GeV}) and \SI{160}{\GeV}.
In the absence of other new physics, measurements of the process $b \to s \gamma$ exclude in Type-II and Type-Y models
a charged Higgs boson with mass below \SI{580}{\GeV} independent of $\tan{\beta}$~\cite{Misiak:2017bgg}.

\begin{figure}[h]
\centering
\includegraphics{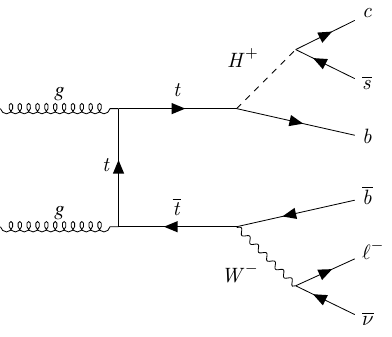} %
\caption{Illustrative Feynman diagram of the signal process.}
\label{Fig:FeynmanSignal}
\end{figure}

This analysis searches for a charged Higgs boson with a mass below the top-quark mass. In this regime the main production mode for charged Higgs bosons is
via $\ttbar$ events, with the charged Higgs boson emerging from rare top-quark decays: $t \rightarrow \Hpm b$.
This analysis focuses on the $\ttbar \rightarrow Wb \Hpm b$ process, with $W \rightarrow \ell\nu$ ($\ell = e, \mu$) and $\Hpm \rightarrow cs$,
resulting in a lepton-plus-jets final state (see Figure~\ref{Fig:FeynmanSignal}). Only events with an electron or muon, including those produced via leptonically decaying $\tau$-leptons,
denoted by $\tau_\mathrm{lep}$, are considered.

This paper is based on $\SI{140}{\ifb}$ of \pp collision data at $\sqrt{s} = \SI{13}{\TeV}$. In contrast to previous $\Hpm \rightarrow cs$ searches, multivariate analysis techniques are deployed here to search for a potential signal in the data.
This approach exploits both the kinematic properties of signal events, including the \Hpm candidate mass, and their flavour composition, which differs from that of the mostly SM \ttbar background events.
The kinematics of the top-quark decays are derived by reconstructing the \ttbar\ event topology.
Flavour-tagging algorithms are utilised to tag jets as $b$-jets, $c$-jets or light-flavour jets.
A dedicated flavour-tagging scheme which facilitates simultaneous tagging of $b$-jets and $c$-jets is adopted.


%
%

%
%
%

\section{ATLAS detector}
\label{Sec:AtlasDetector}

\newcommand{\AtlasCoordFootnote}{%
ATLAS uses a right-handed coordinate system with its origin at the nominal interaction point (IP)
in the centre of the detector and the \(z\)-axis along the beam pipe.
The \(x\)-axis points from the IP to the centre of the LHC ring,
and the \(y\)-axis points upwards.
Polar coordinates \((r,\phi)\) are used in the transverse plane,
\(\phi\) being the azimuthal angle around the \(z\)-axis.
The pseudorapidity is defined in terms of the polar angle \(\theta\) as \(\eta = -\ln \tan(\theta/2)\) and is equal to the rapidity
$ y = (1/2) \ln \left[ (E + p_z)/(E - p_z) \right]$ in the relativistic limit.
Angular distance is measured in units of \(\Delta R \equiv \sqrt{(\Delta y)^{2} + (\Delta\phi)^{2}}\).}

The ATLAS detector~\cite{PERF-2007-01} at the LHC covers nearly the entire solid angle around the collision point.\footnote{\AtlasCoordFootnote}
It consists of an inner tracking detector surrounded by a thin superconducting solenoid, electromagnetic (ECAL) and hadronic calorimeters,
and a muon spectrometer (MS) incorporating three large superconducting air-core toroidal magnets.

The inner-detector system (ID) is immersed in a \qty{2}{\tesla} axial magnetic field
and provides charged-particle tracking in the range \(|\eta| < 2.5\).
The high-granularity silicon pixel detector covers the vertex region and typically provides four measurements per track,
the first hit generally being in the insertable B-layer (IBL) installed before Run~2~\cite{ATLAS-TDR-19,PIX-2018-001}.
It is followed by the SemiConductor Tracker (SCT), which usually provides eight measurements per track.
These silicon detectors are complemented by the transition radiation tracker (TRT),
which enables radially extended track reconstruction up to \(|\eta| = 2.0\).
The TRT also provides electron identification information
based on the fraction of hits (typically 30 in total) above a higher energy-deposit threshold corresponding to transition radiation.

The calorimeter system covers the pseudorapidity range \(|\eta| < 4.9\).
Within the region \(|\eta|< 3.2\), electromagnetic calorimetry is provided by barrel and
endcap high-granularity lead/liquid-argon (LAr) calorimeters,
with an additional thin LAr presampler covering \(|\eta| < 1.8\)
to correct for energy loss in material upstream of the calorimeters.
Hadronic calorimetry is provided by the steel/scintillator-tile calorimeter,
segmented into three barrel structures within \(|\eta| < 1.7\), and two copper/LAr hadronic endcap calorimeters.
The solid angle coverage is completed with forward copper/LAr and tungsten/LAr calorimeter modules
optimised for electromagnetic and hadronic energy measurements respectively.

The MS comprises separate trigger and
high-precision tracking chambers measuring the deflection of muons in a magnetic field generated by the superconducting air-core toroidal magnets.
The field integral of the toroids ranges between \num{2.0} and \qty{6.0}{\tesla\metre}
across most of the spectrometer.
Three layers of precision chambers, each consisting of layers of monitored drift tubes, cover the region \(|\eta| < 2.7\),
complemented by cathode-strip chambers in the forward region, where the background is highest.
The muon trigger system covers the range \(|\eta| < 2.4\) with resistive-plate chambers in the barrel, and thin-gap chambers in the endcap regions.

The luminosity is measured mainly by the LUCID-2~\cite{LUCID2} detector that records Cherenkov light produced in the quartz windows of photomultipliers located close to the beam pipe.

Events are selected by the first-level trigger system implemented in custom hardware,
followed by selections made by algorithms implemented in software in the high-level trigger~\cite{TRIG-2016-01}.
The first-level trigger accepts events from the \qty{40}{\MHz} bunch crossings at a rate below \qty{100}{\kHz},
which the high-level trigger further reduces in order to record complete events to disk at about \qty{1}{\kHz}.

A software suite~\cite{SOFT-2022-02} is used in data simulation, in the reconstruction
and analysis of real and simulated data, in detector operations, and in the trigger and data acquisition
systems of the experiment.


%
%
%

%
%
%

\section{Data and simulated event samples}
\label{Sec:DataMcSamples}

This search is based on data collected from \pp collisions at the ATLAS experiment during LHC \RunTwo at a centre-of-mass energy of $\sqrt{s} = \SI{13}{\TeV}$.
After applying quality requirements, the dataset corresponds to an integrated luminosity of
$140.1 \pm 1.2\,\ifb$~\cite{LUCID2,DAPR-2021-01}.
Signal and background processes, except multijet processes, were simulated using Monte Carlo (MC) event generators.

The main background in this search is SM \ttbar production.
The production of \ttbar and single-top-quark events in the $t\Wboson$-, $s$- and $t$-channels
was modelled with the \POWHEGBOX[v2]~\cite{Frixione:2007nw,Nason:2004rx,Frixione:2007vw,Alioli:2010xd}
generator at next-to-leading order (NLO), using the five-flavour scheme
(four-flavour scheme for single-top-quark $t$-channel events) with the \NNPDF[3.0nlo]~\cite{Ball:2014uwa}
parton distribution function~(PDF) set and the \hdamp parameter\footnote{The
\hdamp parameter is a resummation damping factor and one of the
parameters that controls the matching of \POWHEG matrix elements to
the parton shower and thus effectively regulates the
high-\pt radiation against which the \ttbar system recoils.} set
to 1.5\,\mtop~\cite{ATL-PHYS-PUB-2016-020}.
The events were interfaced to \PYTHIA[8.230]~\cite{Sjostrand:2014zea} to model the parton shower (PS),
hadronisation, and underlying event. For all samples in this search, \PYTHIA[8] used the \NNPDF[2.3lo] set of PDFs~\cite{Ball:2012cx}, and its parameter values were set to those of the A14 tune~\cite{ATL-PHYS-PUB-2014-021}.
The decays of bottom and charm hadrons were performed by \EVTGEN[1.6.0]~\cite{Lange:2001uf}.
To assess the uncertainty in the matching of NLO matrix elements (ME) to the
PS, the nominal samples were compared with samples of \ttbar and single-top-quark events
generated with \MGNLO[2.6.0] and~2.6.2~\cite{Alwall:2014hca} respectively,
using the \NNPDF[3.0nlo] set of PDFs, and interfaced with \PYTHIA[8.230].
The impact of using a different PS and hadronisation model was evaluated
by comparing the nominal samples with alternative samples produced with the
\POWHEGBOX[v2] generator using the \NNPDF[3.0nlo] PDF set and
interfaced with \HERWIG~\cite{Bahr:2008pv,Bellm:2015jjp}.
\HERWIG[7.13] and \HERWIG[7.16] were used for \ttbar\ and single-top-quark events, respectively,
and both used the \HERWIG[7.1] default set of tuned parameters~\cite{Bellm:2015jjp,Bellm:2017jjp}
and the \MMHT[lo] PDF set~\cite{Harland-Lang:2014zoa}.
The $\ttbar$--$t\Wboson$ interference was handled using the diagram removal scheme~\cite{Frixione:2008yi}.
The uncertainty associated with this choice is estimated by comparing the nominal sample with an alternative sample
generated using the diagram subtraction scheme~\cite{Frixione:2008yi,ATL-PHYS-PUB-2016-020}.
The \ttbar production cross-section is calculated at next-to-next-to-leading-order and next-to-next-to-leading-logarithm (NNLO+NNLL) accuracy~\cite{Czakon:2011xx}.
The cross-sections for the three single-top-quark production channels are calculated at NLO~\cite{1103.2792,1001.5034,Kidonakis:2010ux}.
Simulated \ttbar events are categorised according to the flavour of additional jets in the event, using the procedure described in Ref.~\cite{HIGG-2013-27}.
Events with at least one additional $b$-flavour or $c$-flavour jet are labelled as \ttbarHF (where HF stands for \enquote{heavy-flavour}).
The remaining events are labelled as \ttbarLF (where LF stands for \enquote{light-flavour}).
This category is split into \ttbarll and \ttbarlc subcategories according to the decay of the $\PW$ bosons.
If at least one $\PW$ boson decays as $\PW \to cs$ the event falls into the \ttbarlc category, otherwise it falls into the \ttbarll category.
This categorisation is motivated by the fact that the final state is identical for \ttbarlc and signal events.

The rare top-quark processes considered in this analysis are \ttH, \ttW, \ttZ, \tttt, $\ttbar t$, $tHjb$, $tWH$, \tWZ and \tZq.
The \ttH\ events were modelled with the same generators and versions as the exclusive \ttbar events.
The \ttW, \ttZ, \tWZ and \tttt processes were modelled using \MGNLO[2.3.3], the $\ttbar t$ process using \MGNLO[2.2.2],
and the $tHjb$ and $tWH$ processes using \MGNLO[2.6.2], in all cases at NLO with the \NNPDF[3.0nlo] PDF (\NNPDF[3.1nlo] for the \tttt process).
The $\ttbar t$ and \tZq processes were modelled with \MADGRAPH[2.2.2] at LO with the \NNPDF[2.3nlo] PDF.
The events were then interfaced with \PYTHIA[8.186--8.235] and the decays of bottom and charm
hadrons were simulated using the \EVTGEN[1.2.0--1.6.0] program.
The \ttH, \ttW and \ttZ samples were normalised using cross-sections calculated at NLO QCD and NLO EW accuracy using
\MGNLO as reported in Ref.~\cite{deFlorian:2016spz}.

The production of a $W$ or $Z/\gamma^{*}$ boson in association with jets ($V$\;\!+\,jets) and of dibosons ($VV$) was modelled with the
\SHERPA[2.2.11]~\cite{Bothmann:2019yzt} and \SHERPA[2.2.1] generators, respectively, for both the ME and PS.
The only exception is the $VV \to \ell \nu\nu\nu$ process, which was modelled with \SHERPA[2.2.2].
The NLO ME for up to two partons (one parton) and leading-order ME
for up to five (three) partons were calculated with the \textsc{Comix}~\cite{Gleisberg:2008fv}
and \OPENLOOPS~\cite{Buccioni:2019sur,Cascioli:2011va,Denner:2016kdg} libraries
for $V$\;\!+\,jets ($VV$) events.
They were matched with the \SHERPA PS~\cite{Schumann:2007mg} using the \MEPSatNLO
prescription~\cite{Hoeche:2011fd,Hoeche:2012yf,Catani:2001cc,Hoeche:2009rj}
and the set of tuned parameters developed by the \SHERPA authors.
The \NNPDF[3.0nnlo] set of PDFs was used.
Signal events were modelled by first generating top-quark pairs, similar to the SM \ttbar background, using the \POWHEGBOX[v2]
generator at NLO with the \NNPDF[3.0nlo] PDF set and the \hdamp parameter set to 1.5\,\mtop.
The decays $t \to \Hpm \Pbottom$  and $t \to \Wpm\Pbottom$ were modelled by \MADSPIN~\cite{Artoisenet:2012st}
using the Type-II 2HDM~\cite{Branco:2011iw,Alloul:2013bka} for BSM decays.
Subsequent decays of the $\Hpm$ and $\Wpm$ bosons, as well as the showering of the final-state hadrons, were modelled by \PYTHIA[8.307].
The $\Wpm$ were forced to decay leptonically, with all three lepton flavours allowed.
The $\Hpm$ were forced to decay into a $cs$-quark pair.
The decays of bottom and charm hadrons were performed by \EVTGEN[1.7.0].
Signal samples were generated with zero decay width for twelve charged-Higgs-boson mass points:
eleven in steps of \SI{10}{\GeV} from \SI{60}{\GeV} to \SI{160}{\GeV}, and one at \SI{168}{\GeV}.
The signal samples are denoted by $\Hpm_{x}$, where $x$ is the mass of the charged Higgs boson in \GeV.
If the mass difference between the \Hpm and $W$ bosons is smaller than one of their total widths, the interference term might
be of the order of a few percent of the \Hpm contribution. The size and the sign of the interference term depend on the
model~\cite{MoosaviNejad:2019agw}. For larger mass differences, the interference term
can be omitted with high accuracy. The interference term is neglected in this analysis for all mass points.
Production of charged Higgs bosons via single-top-quark processes is neglected in this analysis because
such events usually do not contain a prompt lepton and are therefore suppressed by the event selection (cf.\ Section~\ref{Sec:ObjectDefinition}),
and also because the production cross-section is much smaller than for \ttbar processes.

The effect of multiple interactions in the same and neighbouring bunch
crossings (\pileup) was modelled by overlaying the simulated hard-scattering event with
inelastic \pp events generated with \PYTHIA[8.186]~\cite{Sjostrand:2007gs}
using the \NNPDF[2.3lo] PDF set and the A3 set of tuned parameters~\cite{ATL-PHYS-PUB-2016-017}.
Events in the nominal background samples were passed through the full ATLAS detector simulation~\cite{SOFT-2010-01} based on \GEANT~\cite{Agostinelli:2002hh}.
Signal, $\ttbar\ttbar$, $tH$ and alternative samples were passed through a fast simulation in which the response of the calorimeter is parameterised~\cite{ATL-PHYS-PUB-2010-013}.
Both simulation methods were found to provide a similar level of modelling accuracy for the physics objects used in the analysis.
A full list of samples used in this search is summarised in Table~\ref{Tab:Samples}.

\begin{table}
\centering
\caption{Generators used to simulate the signal and background processes.
The symbol $q$ is used for $u,d,c,s$ quarks.
For the signal processes the subscript \enquote{$x$} is a placeholder for the mass of the charged Higgs boson in \GeV.
For the \enquote{Other top} and $VV$ processes, only the range of used generator versions is quoted.
The exact generator version used for each process is described in Section~\ref{Sec:DataMcSamples}.}
\resizebox{\textwidth}{!}{
\begin{tabular}{l l l l}
\toprule
Name & Process & ME Generator & PS and hadronisation\\
\midrule
\textbf{Signal} & & & \\
$\quad \Hpm_{x}$ & $\ttbar \to \Hpm(\to cs) W^{\mp}(\to \ell \nu_{\ell}) b\bar{b}$ & \POWHEGBOX[v2] & \MADSPIN+ \PYTHIA[8.307]+ \EVTGEN[1.7.0] \\
\hline
\textbf{Top-quark} & & & \\
$\quad$ \ttbarll & $\ttbar \to W^{\pm}(\to \ell \nu_{\ell}) W^{\mp}(\to ud) b\bar{b}$ & \multirow{7}{*}{\POWHEGBOX[v2]} & \multirow{8}{*}{\PYTHIA[8.230]+ \EVTGEN[1.6.0]} \\
$\quad$ \ttbarlc & $\ttbar \to W^{\pm}(\to \ell \nu_{\ell}) W^{\mp}(\to cs) b\bar{b}$ & & \\
$\quad$ \ttbarHF & $\ttbar \to W^{\pm}(\to \ell \nu_{\ell}) W^{\mp}(\to q\bar{q}) b\bar{b}\,{+}\,{\geq}1 c/b$ & & \\
$\quad$ $\ttbar(\text{allHad})$ & $\ttbar \to W^{\pm}(\to q\bar{q}) W^{\mp}(\to q\bar{q})$ & & \\
$\quad$ \singleTopWt& $tW$ & & \\
$\quad$ \singleTop & single $t$-quark $s$- \& $t$-channel & & \\
$\quad$ $\ttbar H$ & $\ttH$ & & \\
$\quad$ \multirow{2}{*}{\OtherTop} & $\ttW$, $\ttZ$, $\tttt$, $tHjb$, $tWH$, $tWZ$, & \MGNLO[2.3.3--2.6.2] & \multirow{2}{*}{\PYTHIA[8.186--8.230]+ \EVTGEN[1.2.0--1.6.0]}\\
& $\ttbar t$, $\tZq$ & \MADGRAPH[2.2.2--2.3.3] & \\
\hline
\textbf{Weak-boson} & & & \\
$\quad$ \Wjets & \Wjets & \multirow{2}{*}{\SHERPA[2.2.11]} & \multirow{2}{*}{\SHERPA[2.2.11]}\\
$\quad$ \Zjets & \Zjets & & \\
$\quad$ $VV$ & $WW$, $WZ$, $ZZ$ & \SHERPA[2.2.1--2.2.2] & \SHERPA[2.2.1-- 2.2.2] \\
\bottomrule
\end{tabular}
}
\label{Tab:Samples}
\end{table}


%
%
%
%

%
%
%

\section{Object definition and event selection}
\label{Sec:ObjectDefinition}

Tracks are required to have transverse momentum (\pT) greater than $\SI{500}{\MeV}$, $|\eta| < 2.5$, and at
least seven hits in the pixel and SCT detectors.
A maximum of one (two) of the expected hits may be missing from the pixel (SCT) detector,
and no more than one hit may be shared with other tracks~\cite{PERF-2015-08}.
Events are required to have at least one primary vertex reconstructed from two or more associated tracks~\cite{ATL-PHYS-PUB-2015-026}.
If multiple vertices are found, the one with the highest scalar sum of the $\pT^2$ of associated tracks is selected as the primary vertex.

Electrons are reconstructed from topological energy clusters in the ECAL that are matched to tracks in the ID~\cite{EGAM-2018-01}.
Electrons are required to have $\pt > \SI{10}{\GeV}$ and $|\eta| < 2.47$,
excluding the barrel--endcap transition region $1.37 < |\eta| < 1.52$. They must pass track-quality requirements
followed by a loose likelihood-based selection that requires the shower profile to be compatible with that of the electromagnetic
shower. Electrons are required to have transverse ($d_0$) and longitudinal ($z_0$) impact parameters,
measured relative to the beam-line and primary vertex respectively, satisfying $|d_0|/\sigma (d_0)<5$ and $|z_0 \sin{\theta}| < \SI{0.5}{\mm}$.
Isolation requirements are applied via a boosted decision tree (BDT) which was trained on track-isolation, cluster-isolation,
and secondary-vertex information, referred to as \enquote{non-prompt-lepton BDT}~\cite{HIGG-2017-02,EGAM-2019-01}. The electron energy scale and resolution calibrations are obtained from
$Z\rightarrow ee$ events and applied to data and simulations, respectively~\cite{EGAM-2018-01}.

Muon candidates are reconstructed by matching MS tracks to ID tracks. In the absence of full tracks in the MS, muons can be
reconstructed from ID tracks extrapolated to the MS which match at least three loosely aligned MS hits. The information from the ID
and the MS, and the energy loss in the calorimeters, are then used in a combined track fit~\cite{MUON-2018-03}.
Muons have to satisfy $\pt > \SI{10}{\GeV}$ and $|\eta| < 2.5$, and pass quality requirements based on the number of hits
used to reconstruct the tracks. Muons are also required to satisfy $|d_0|/\sigma (d_0)<3$ and $|z_0 \sin{\theta}| < \SI{0.5}{\mm}$.
Lastly, isolation requirements are also made based on the non-prompt-lepton BDT.

Jets are reconstructed with the \akt jet clustering algorithm~\cite{Cacciari:2008gp,Fastjet} with a radius parameter $R=0.4$. The clustering is applied to
noise-suppressed positive-energy topological energy clusters~\cite{PERF-2014-07,ATL-PHYS-PUB-2015-036} and charged-particle tracks, processed
using a particle-flow algorithm~\cite{PERF-2015-09}. Jet energies are corrected for contributions from \pileup, calibrated using energy- and $\eta$-dependent
correction factors determined from comparisons between particle-level objects and reconstructed physics objects in simulated events, and then corrections are
applied to account for effects due to the initiating-parton type and hadron composition~\cite{JETM-2018-05}. In data, a residual in situ correction is applied in
order to correct for differences relative to simulation. Jets in the analysis are required to have $\pt>\SI{25}{\GeV}$ and $|\eta|<2.5$. Jets with $\pt<\SI{60}{\GeV}$
and $|\eta|<2.4$ also have to pass a jet-vertex-tagger~\cite{PERF-2014-03} requirement to reduce the number of selected jets which originate from \pileup.

Jets containing $b$- or $c$-hadrons are identified with the DL1r tagger~\cite{FTAG-2019-07}, which is a multivariate
classification algorithm based on a deep neural network using information about the impact parameters of tracks, the jet kinematics, and displaced vertices.
The $b$- and $c$-tagging scores are based on log-likelihood ratios of the neural-network output scores.
To assign jets to top quarks or to \Hpm boson candidates,
one needs to identify $b$- and  $c$-quark-initiated jets simultaneously and distinguish them from the light-flavour jets.
Correction factors are applied to the simulated events to compensate for differences between data and simulation in the $b$- and $c$-tagging efficiencies
or misidentification rates for $b$-jets, $c$-jets and light-flavour jets~\cite{FTAG-2018-01,FTAG-2020-08,FTAG-2019-02}.
This search uses a pseudo-continuous flavour-tagging (PCFT) calibration  with five exclusive calibrated bins.
Jets passing a fixed $b$-tagging working point (WP) defined by $b$-jet efficiencies, measured in \ttbar events, of \SI{70}{\%} and \SI{60}{\%} have PCFT scores of 3 and 4, respectively.
These two $b$-tagging WPs have background tagging efficiencies for $c$-jets (light-jets) of \SI{7.9}{\%} (\SI{0.18}{\%}) and \SI{2.7}{\%} (\SI{0.05}{\%}), respectively~\cite{HIGG-2020-20}.
Other jets ($b$-veto) receive a PCFT score of 1 or 2 if they pass a fixed $c$-tagging score defined by a $c$-jet efficiency of \SI{45}{\%} or \SI{24}{\%}, respectively.
These two $c$-tagging WPs have background tagging efficiencies for $b$-jets (light-jets) of \SI{16.3}{\%} (\SI{7.4}{\%}) and \SI{4.8}{\%} (\SI{0.9}{\%}), respectively~\cite{HIGG-2020-20}.
Jets passing none of the $b$- and $c$-tagging WPs (untagged) are assigned a PCFT score of 0.
Any jet passing the loosest calibrated $b$($c$)-tagging WP is referred to as a $b$($c$)-tagged jet.

The missing transverse momentum (\metvec) is defined as the negative vector sum of the transverse momenta of all reconstructed and calibrated leptons and jets,
and all tracks matched to the primary vertex but not to other reconstructed objects in the event~\cite{JETM-2020-03}.
The absolute value of \metvec\ is denoted by \MET.

An overlap-removal procedure is applied to resolve ambiguities where multiple physical objects are reconstructed
from the same detector signature. The angular distance $\DeltaR$ is used to measure the overlap of two reconstructed objects.
The following procedure is applied in order:
\begin{enumerate}
\item any calorimeter-tagged muon~\cite{MUON-2018-03} sharing a track with an electron is removed;
\item any electron sharing a track with a muon is removed;
\item any jet within $\DeltaR=0.2$ of an electron is removed;
\item any electron within $\DeltaR=0.4$ of a jet is removed;
\item any jet with less than 3 tracks that is within $\DeltaR=0.2$ of a muon is removed;
\item any jet with less than 3 tracks that has a muon ID track ghost-associated~\cite{Cacciari:2008gp,Cacciari:2008gn} with it is removed;
\item any muon within $\DeltaR=0.4$ of a jet is removed.
\end{enumerate}

Events were recorded with a single-electron or single-muon trigger with a threshold requirement imposed on the lepton \pT. For the data-taking periods 2015 and 2016--2018, the lowest electron \pT threshold was \SI{24}{\GeV}
or \SI{26}{\GeV} respectively, and similarly the lowest muon-\pt threshold was \SI{20}{\GeV} or \SI{26}{\GeV}. The trigger includes lepton identification and isolation requirements
based on ID or ECAL measurements~\cite{TRIG-2018-01,TRIG-2018-05,TRIG-2019-04}.
Furthermore, events are required to have exactly one offline reconstructed lepton with $\pt>\SI{27}{\GeV}$ that meets the \enquote{medium} identification and isolation criteria~\cite{EGAM-2018-01, MUON-2018-03}.
Events with an additional lepton with $\pt>\SI{10}{\GeV}$ that satisfies the medium identification criteria are vetoed to reduce dileptonic \ttbar\ and \Zjets\ backgrounds.
The offline reconstructed lepton is required to be geometrically matched ($\Delta R<0.1$) to the online reconstructed lepton which fired the trigger.
At least four jets with $\pt>\SI{25}{\GeV}$ have to be present, and at least one is required to be $b$-tagged.
Finally, events in the signal region are required to have exactly one \enquote{tight} identified and isolated lepton and at least two $b$-tagged jets~\cite{EGAM-2018-01, MUON-2018-03}.
The \ttbar\ background contribution in the signal region is about \SI{92}{\%}.


%
%
%
%
%
%
%

%
%
%

\section{Background modelling}
\label{Sec:BackgroundModelling}

The main background in this search is \ttbar production in association with jets.
As in many other analyses targeting a kinematic phase space similar to the signal region of this search,
differences between MC-based background predictions and data are observed in multiple kinematic quantities~\cite{TOPQ-2020-02,HDBS-2019-24}.
This disagreement can be attributed to missing higher-order QCD and electroweak corrections in \ttbar MC simulation leading to harder
top-quark \pt spectra in simulation than in data~\cite{Czakon:2017wor}.
A data-driven correction is derived to improve the modelling of the \ttbar background and signal, particularly of
\pT-dependent variables.

The correction is derived as a function of \ST.
The \ST\ variable is defined as the sum of the scalar transverse momenta of all calibrated objects in the event,
i.e.\ jets, leptons and \MET, and is therefore related to the transverse momenta of individual top quarks.
The corrections are derived in bins of the number of jets in the event ($\Njets=4,5,6,7,8,{\geq}9$),
as this quantity also shows discrepancies between MC events and data and is correlated with \ST.
The correction is derived in the signal region, since the \ttbar contribution is around \SI{92}{\%} in that region, and is
applied to signal events as well as \ttbar events because the mismodelling is expected to affect the signal MC prediction in the same way.
It was checked that the \ST\ distribution in bins of \Njets\ is similar for \ttbar and signal events.
A possible signal contribution in data will therefore not change the correction weights and the correction will not bias the signal extraction.

The \ttbar correction weights are defined as the ratio of data templates (with non-\ttbar backgrounds subtracted) to MC \ttbar templates.
In order to mitigate the effects of statistical fluctuations in the data and simulation samples, a linear + exponential function is fitted to the derived \ttbar correction weights.
The fit is performed separately for even and odd event numbers to avoid overfitting. The weights from the fit to even-numbered events are applied to odd-numbered events and vice versa.

Figure~\ref{Fig:TtbarCorr} shows the \ST\ and lepton-\pt distributions after applying the \ttbar correction. The red dashed line represents the total background prediction before applying the correction. Agreement between the data and MC prediction clearly improves for $\ST$ and other related distributions, especially at high values.
The \ST\ and lepton-\pt distributions of the alternative \ttbar MC samples differ significantly from those of the nominal \ttbar sample.
Hence, alternative \ttbar correction weights are derived and
applied to these events.
\begin{figure}
\centering
\subfloat[]{\includegraphics[width=0.49\textwidth]{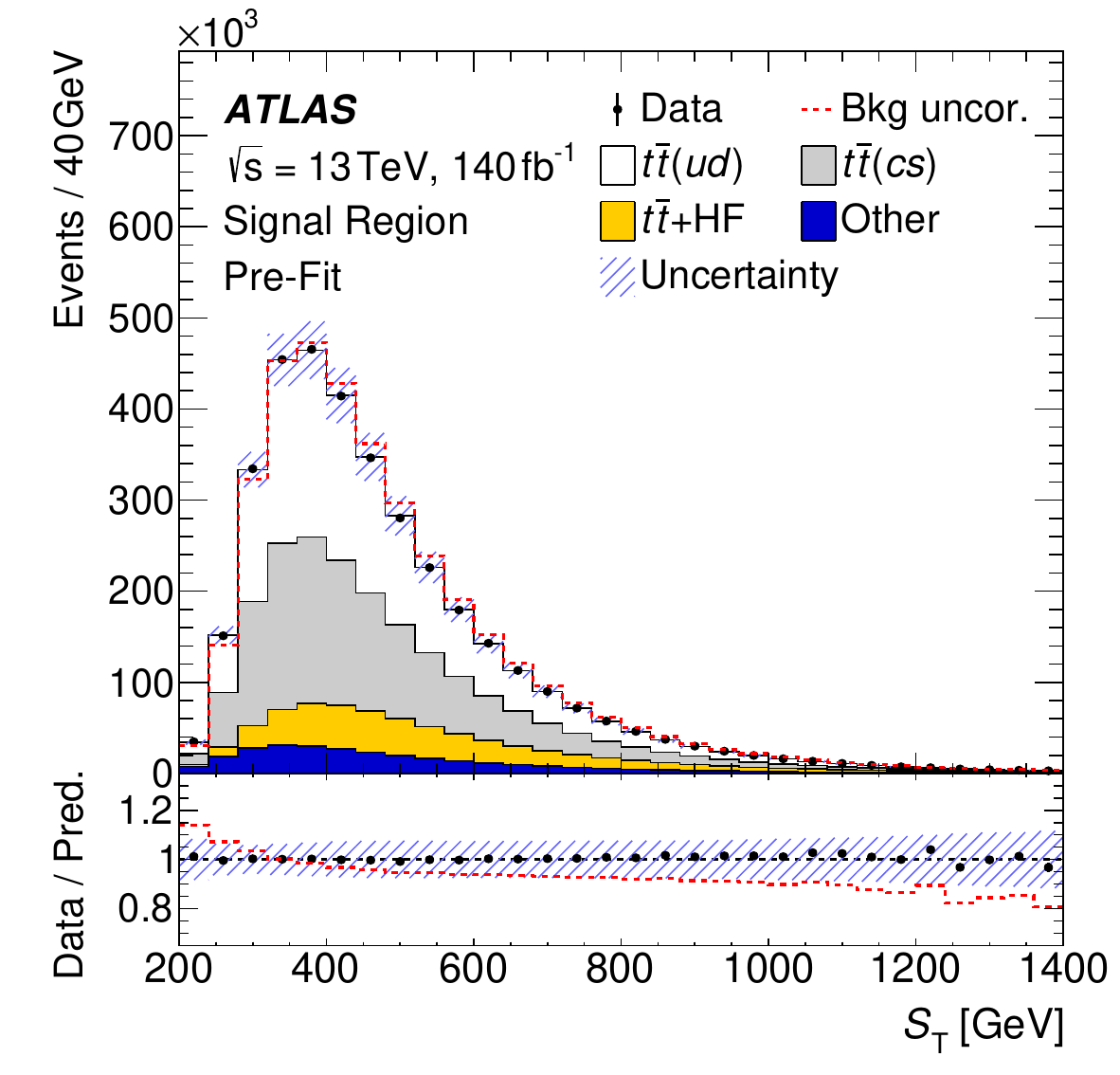}}
\subfloat[]{\includegraphics[width=0.49\textwidth]{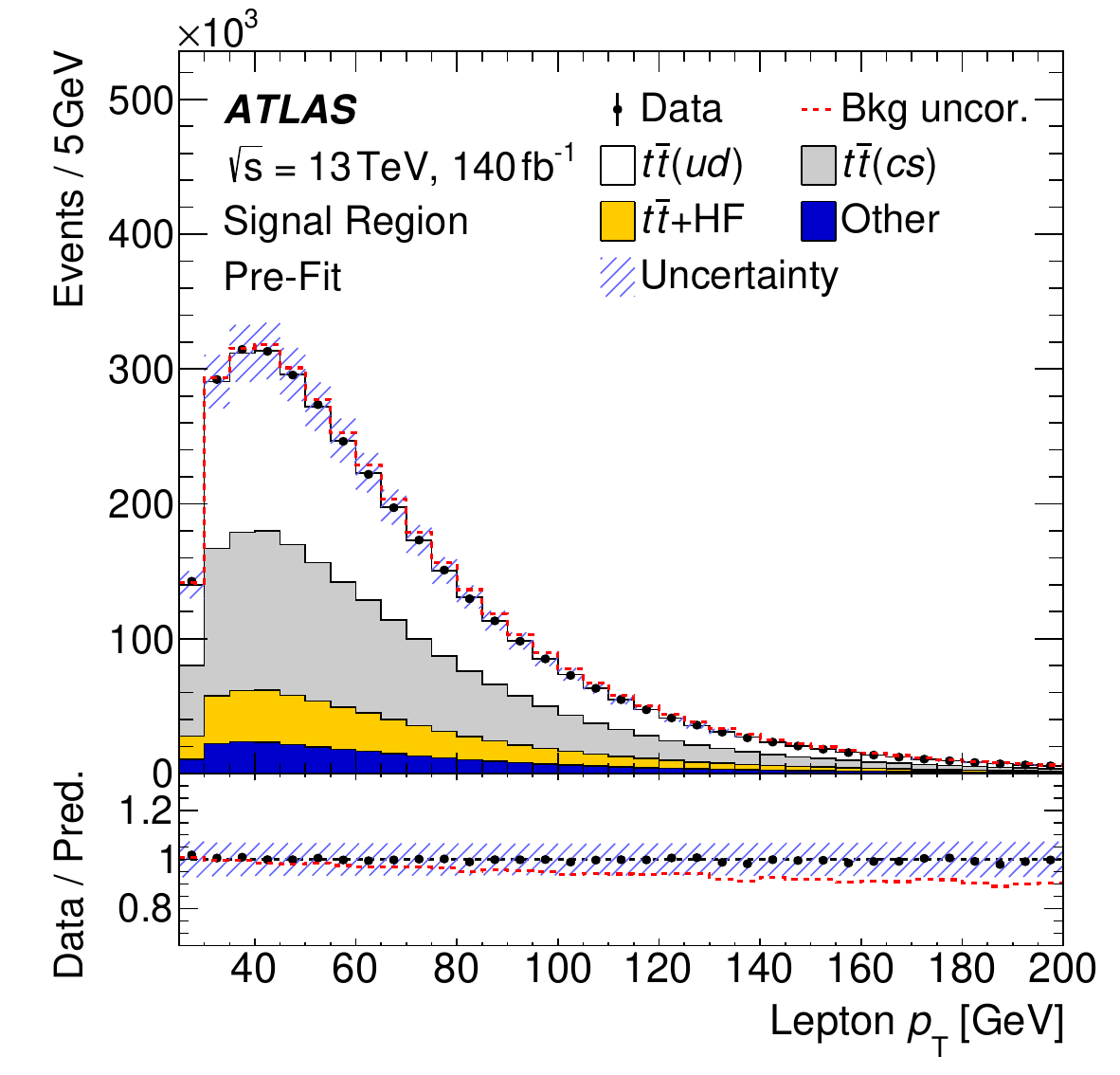}}
\caption{Distribution of (a) \ST\ and (b) lepton \pt after applying the \ttbar correction.
The processes $\ttbar(\text{allHad})$, \singleTopWt, \singleTop, $\ttbar H$, \OtherTop, \Wjets, \Zjets, and $VV$
listed in Table~\ref{Tab:Samples} are combined with the multijet background in the \enquote{Other} category.
The uncertainty band represents the combined statistical and systematic uncertainty of the prediction.
The red dashed line represents the total background prediction before applying the correction.}
\label{Fig:TtbarCorr}
\end{figure}

Multijet (MJ) processes can contribute to the background when jets are misidentified as leptons or when real
non-prompt leptons are produced in the decays of heavy-flavour hadrons.
A data-driven method commonly called the ABCD method (see e.g.\ Ref.~\cite{CDF:1990kbl}) is used to estimate the MJ background in the signal region.
The lepton isolation requirement and the number of $b$-tagged jets are used to define the four ABCD regions.
The shape of the MJ background is estimated from a region with a looser lepton-isolation requirement.
The normalisation is derived from events with exactly one $b$-tagged jet.
The ABCD method is applied separately to electron and muon events.
The size of the total MJ background in the signal region is about \SI{0.3}{\%} of the total estimated background.\footnote{%
The MJ background is considered when deriving the \ttbar correction.
The \ttbar correction weights are also applied in the MJ-enriched regions.
The reciprocal dependence of the MJ background estimate and the \ttbar correction is considered when deriving the correction weights.
The impact on the correction weights is minor due to the overall small MJ contribution.
}


%
%
%
%
%
%
%
%
%

%
%
%

\section{Analysis strategy}
\label{Sec:AnalysisStrategy}

The presence of a potential signal in data is quantified first by reconstructing the \ttbar event topology, and then extracting the signal using a BDT.
The \ttbar-system is reconstructed by using calibrated physics objects, i.e.\ leptons, \metvec\ and jets, as proxies for the lepton, neutrino and quarks from the top-quark decays. Kinematic properties of the reconstructed \ttbar-system are then used to train a BDT to classify events as signal or background.

\subsection{\ttbar-system reconstruction}
\label{Sec:EventReconstruction}

The \ttbar-system consists of a semileptonically decaying top quark (\tlep) and a hadronically decaying top quark (\thad).
The \tlep\ decays into a $b$-quark (\blep) and a $W$ boson, which decays into a lepton and neutrino.
The \thad\ decays into a $b$-quark (\bhad) and a $\Hpm$ or $W$ boson, which decays into  $c$- and $s$-quarks or other quarks
(\ljet1, \ljet2), respectively.

The lepton from the $W$-boson decay is unambiguously identified as the single reconstructed lepton. The momentum of the neutrino from the $W$-boson decay is
reconstructed using \metvec\ and a $W$-boson mass constraint. The neutrino pseudorapidity, $\eta_{\nu}$, is calculated by setting the invariant
mass of the lepton and neutrino equal to the $W$-boson mass, $m_{W}=\SI{80.379}{\GeV}$~\cite{Zyla:2020zbs}:
\begin{equation*}
\label{Eq:NuEta}
\eta_{\nu} = \eta_{\ell} \pm \mathrm{arccosh} \left( \frac{m_{W}^2}{2 \pt^{\nu} \pt^{\ell}} + \cos \left(\phi_{\nu} - \phi_{\ell} \right) \right).
\end{equation*}
This equation generally has two solutions and the one that is chosen depends on the jet labelling, described below.
If the argument of arccosh is exactly one, there is only one solution ($\eta_{\nu} = \eta_{\ell}$).
Due to reconstruction inefficiencies or additional neutrinos in the event, e.g.\ from the decay of $\tau_\mathrm{lep}$,
it may also happen that the argument of arccosh is smaller than one, for which arccosh is not defined.
In this case the argument of arccosh is set to one.
This means $\eta_{\nu} = \eta_{\ell}$, in which case the invariant mass of the lepton and the neutrino exceeds $m_{W}$.
The latter case occurs for about \SI{35}{\%} of the simulated \ttbar events.

The labelling of the jets as \blep, \bhad, \ljet1 and \ljet2 suffers from a combinatorics problem,
which the analysis tries to resolve by comparing the top-quark candidate's mass with the predicted top-quark mass.
Because the mass resolution differs between semileptonically and hadronically decaying top quarks, the approach adopted uses
probability density functions of the reconstructed top-quark masses ($\mathrm{PDF}_t$).
These are built from \ttbar MC events, using reconstructed jets matched to the generator-level (\enquote{truth}) quarks. Since there is only one lepton candidate, no matching is applied for it.
For the neutrino, if there are two solutions for $\eta_{\nu}$, the one closest to the true value of $\eta_{\nu}$ is selected.
The \enquote{truth} quarks are geometrically matched to the closest reconstructed jet within $\DeltaR=0.4$.
The $b$- and $c$-quarks are only matched if the reconstructed jet has a respective \enquote{truth} hadron ($\pT>\SI{5}{\GeV}$) associated with it.
Other quark types are only matched to the reconstructed jet if no heavy-flavour hadron is associated with it.
If multiple \enquote{truth} quarks are matched to the same reconstructed jet, ambiguities are resolved by minimising the sum of the (four) $\DeltaR$ values between the \enquote{truth} quarks and
any reconstructed jet within $\DeltaR=0.4$ of the \enquote{truth} quarks. In roughly \SI{53}{\%} of the events, at least one \enquote{truth} quark cannot be matched to a reconstructed jet.
Such cases are typically associated with \enquote{truth} quarks produced outside the detector acceptance.
Such events are not considered in the $\mathrm{PDF}_t$.
In order to get a smooth prediction over the full top-quark mass range, Crystal~Ball~+ Cauchy and Crystal~Ball~+ Gaussian functions~\cite{Gaiser:1982y,Oreglia:1980cs,Skwarnicki:1986xj}
are fitted to the \tlep-mass and \thad-mass $\mathrm{PDF}_t$, respectively.
The $\mathrm{PDF}_t$ are shown with the corresponding fits in Figure~\ref{Fig:TopMassPDFs}.
\begin{figure}
\centering
\subfloat[]{\includegraphics[width=0.49\textwidth]{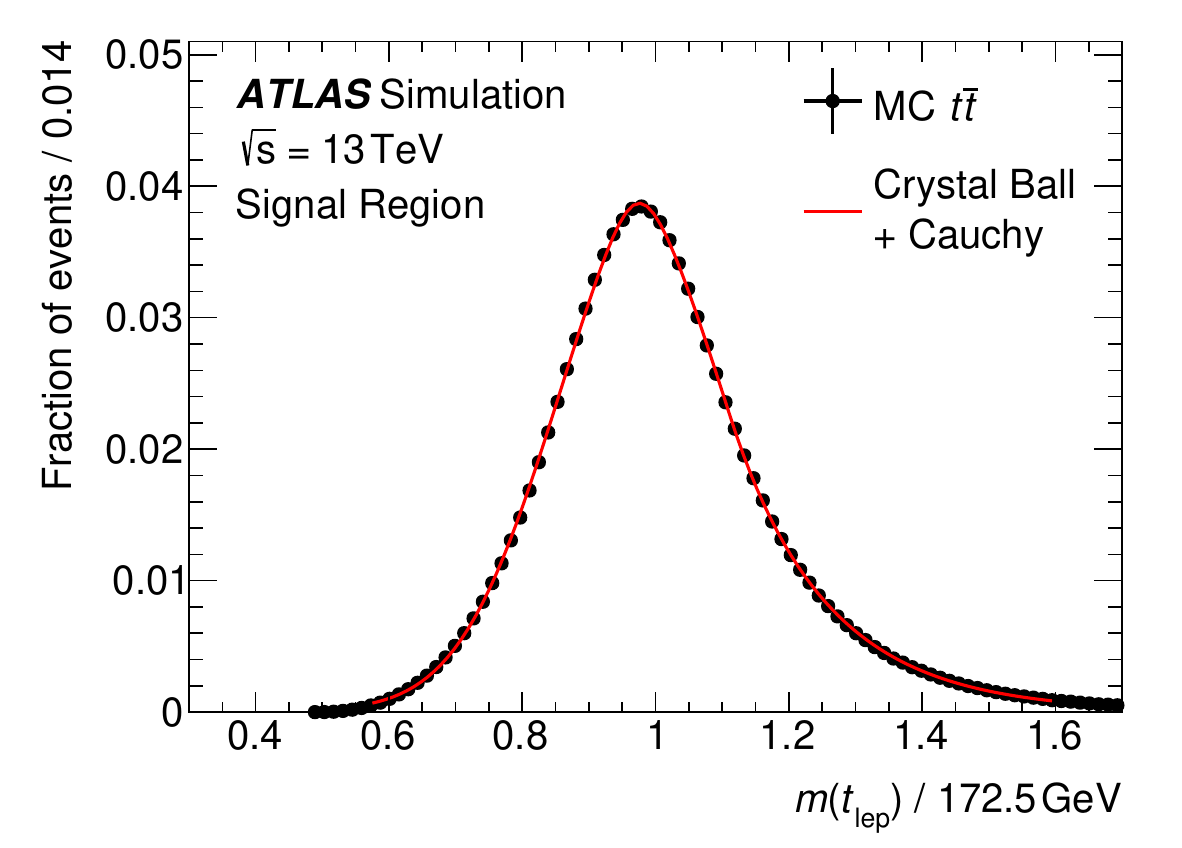}}
\subfloat[]{\includegraphics[width=0.49\textwidth]{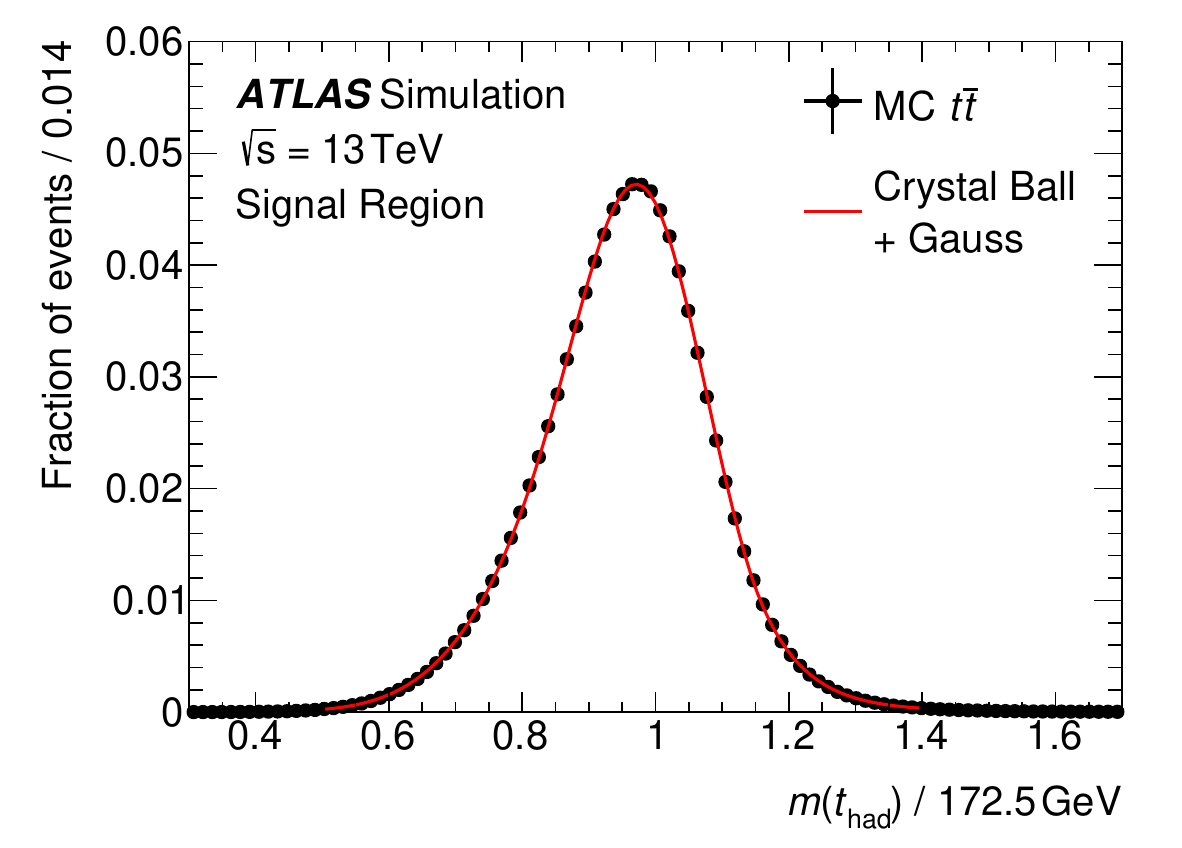}}
\caption{Probability density functions of the reconstructed mass for (a) semileptonically and (b) hadronically decaying top quarks.}
\label{Fig:TopMassPDFs}
\end{figure}

The derived top-quark mass $\mathrm{PDF}_t$ are then used to label jets in an event. All possible permutations of \blep, \bhad, \ljet1, \ljet2 labellings
and $\eta_{\nu}$ solutions are built simultaneously. The permutation with the largest product of the \tlep-mass and \thad-mass $\mathrm{PDF}_t$, i.e.\
$\mathrm{PDF}_{\tlep}(m^{\mathrm{cand}}_{\tlep}) \times \mathrm{PDF}_{\thad}(m^{\mathrm{cand}}_{\thad})$, is chosen, and the jets are labelled accordingly.
The highest $\mathrm{PDF}_t$ product value is denoted by \ProbLabel, and \ProbLabel\ divided by the
sum of $\mathrm{PDF}_t$ product values for all considered permutations is denoted by \NormProbLabel.

However, a few physics-motivated requirements are applied to limit the number of jet permutations.
A maximum of six jets are considered when building the permutations.
The $b$-jets are always considered, whereas the highest-\pt non-$b$-jets are considered first.
The jets labelled as \blep-jets or \bhad-jets have to be $b$-tagged and their PCFT scores are required to be greater than or equal to
the PCFT scores of \ljet1 and \ljet2.
It is also required that the \pt of \ljet1 is greater than the \pt of \ljet2. This requirement removes redundant permutations, since
interchanging \ljet1 and \ljet2 yields identical values for the top-quark and charged-Higgs-boson candidate masses.
Finally, if multiple jets among \bhad, \ljet1, \ljet2 are $b$-tagged and
the $b$-tagged jets have the same PCFT score, multiple permutations will yield the same top-quark candidate mass.
In these rare cases the $b$-tagged jet with the larger \pt is labelled as the \bhad-jet.
In \SI{64}{\%} of the events entering the $\mathrm{PDF}_t$, all four jets are labelled correctly. The performance for signal events is comparable.

\subsection{Multivariate signal extraction}
\label{Sec:MultivariateSignalExtraction}

The multivariate signal extraction exploits differences between the characteristics of the charged Higgs boson and the $W$ boson.
These are the boson mass, spin and decay properties. These differences are seen in many variables
related to flavour-tagging and the kinematics of the top-quark decay products.
The discriminating variables are combined into a single discriminant through the use of BDTs.

The BDT classifies events as signal-like or background-like. Background-like events receive BDT scores close to 0, whereas signal-like events receive
BDT scores close to 1. Separate BDTs are trained for each signal mass-point hypothesis with 5-fold cross-training and using the gradient boosting technique~\cite{hastie_09_elements-of.statistical-learning}.
All simulated background samples listed in Section~\ref{Sec:DataMcSamples} are used in the BDT training.
Events included in the training have to pass the event selection described in Section~\ref{Sec:ObjectDefinition}.
Training and application are carried out within the \xgb~\cite{Chen:2016btl} framework.

Any variable describing the properties of top quarks and their decay products is considered as an input to the BDT training.
The final set of BDT input variables is obtained by recursively removing variables with relatively small power to separate the signal
from the background, and low importance in the BDT training, until a statistically significant loss in performance is observed.
The performance is quantified by the area under the receiver-operating-characteristic curve. This optimisation is performed with the
\SI{130}{\GeV} signal sample. The signal and \ttbar background
kinematics are very similar for the signal mass points close to the $W$-boson mass but differ greatly for signal mass points closer to the
top-quark mass; the intermediate \SI{130}{\GeV} mass point covers both cases.

The final BDT uses the 26 input variables which are listed in Table~\ref{Tab:BDTInputVariables}.
The variables can be sorted into three categories: top-quark kinematic variables, event variables, and flavour-tagging variables.
The first category contains variables related to the kinematics of the
top quarks and their decay products. These variables are mainly sensitive to the mass difference between the $\Hpm$
and $W$ bosons. If the mass difference is comparable to or larger than the dijet mass resolution~\cite{JETM-2018-05},
the kinematic variables exhibit larger separation power than variables in the other categories.
The most important variables (in order of decreasing area under the receiver-operating-characteristic curve)
are the invariant mass of \ljet1 and \ljet2 and the transverse momentum of the \bhad-jet, followed by the invariant mass of \ljet1 and \bhad-jet,
the invariant mass of\ljet2 and \bhad-jet, and the transverse momenta of \ljet1 and \ljet2.
Variables related to the \ttbar-system, the \tlep-quark and the decay products provide separation power between signal and
non-\ttbar backgrounds. In addition, they can carry information about possible wrong jet labelling and add
information via correlation with other input variables.

The $\cos{\theta}$ variable is the only variable that is sensitive to the spin of the boson from the \thad\ decay and therefore
shows separation power for any $\Hpm$ boson mass. The angle $\theta$ is defined as the angle between the \bhad-jet and the up-type-quark-initiated jet
from the hadronically decaying boson ($\Hpm$ or $W$) in the boson's rest frame. The up-type-quark-initiated jet is identified with the help of the PCFT scores of \ljet1 and \ljet2.
The $c$-tagged jets are prioritised over $b$-tagged jets, which in turn are prioritised over untagged jets.
If the PCFT scores of \ljet1 and \ljet2 are identical, \ljet1 is assigned to be the jet from the up-type quark.
The $\cos{\theta}$ distribution is flat for spin-0 particles, like the $\Hpm$ boson. For spin-1 particles, like the $W$ boson, the
distribution is more complex because polarisation effects play a role, and there are fewer events close to $-1$ and 1. However,
it is difficult to identify the quark flavours in the decay of the boson. In addition, the $\cos{\theta}$
distribution is heavily affected by jet resolution effects because three jets are used in the calculation.

\begin{table}[h]
\centering
\caption{Final list of BDT input variables used in the training.}
\begin{tabular}{l | l | l}
\toprule
Variable type            & Variable name & Definition\\
\midrule
\multicolumn{3}{c}{\textbf{Top-quark kinematic variables}}\\
\hline
\multirow{8}{*}{\thad}
& \ljet1 \pt   & \pt of \ljet1-labelled jet \\
& \ljet2 \pt   & \pt of \ljet2-labelled jet \\
& \bhad\ \pt    & \pt of \bhad-jet \\
& \bhadthad\ $p$ & Momentum of \bhad-jet in \thad\ rest frame \\
& dijet mass            & Invariant mass of \ljet1+\ljet2 jets \\
& (\ljet1+\bhad) mass   & Invariant mass of \ljet1+\bhad\ jets \\
& (\ljet2+\bhad) mass   & Invariant mass of \ljet2+\bhad\ jets \\
& $\cos{\theta}$        & Boson spin sensitive variable \\
\hline
\multirow{4}{*}{\tlep}
& \blep\ \pt    & \pt of \blep-jet \\
& Lepton \pt    & \pt of reconstructed lepton \\
& \Wboson mass   & Invariant mass of reconstructed \Wboson boson \\
& \tlep\ mass    & Invariant mass of reconstructed \tlep \\
& \tlep\ \pt     & \pt of reconstructed \tlep \\
\hline
\multirow{2}{*}{\ttbar-system}
& $\Delta R$(\blep,\bhad)   & $\Delta R$ between the \blep-jet and \bhad-jet \\
& \ttbar mass   & Invariant mass of \thad+\tlep \\
\hline
\multicolumn{3}{c}{\textbf{Event variables}}\\
\hline
\multirow{4}{*}{Event level}
& $N_{\text{jets}}$  & Number of jets in the event \\
& \ST  & Scalar \pT sum of all calibrated objects \\
& \makecell[l]{\NormProbLabel} & \makecell[l]{Normalised probability of correct jet labelling} \\
\hline
\multicolumn{3}{c}{\textbf{Flavour-tagging variables}}\\
\hline
\multirow{4}{*}{Flavour-tagging score}
& \ljet1 PCFT  & PCFT score of \ljet1 \\
& \ljet2 PCFT  & PCFT score of \ljet2 \\
& \bhad\ PCFT  & PCFT score of \bhad-jet \\
& \blep\ PCFT  & PCFT score of \blep-jet \\
\hline
\multirow{4}{*}{Number of tags}
& $N_{c\textrm{-tagLo}}$  & Number of jets passing loose $c$-tag WP ($b$-veto) \\
& $N_{c\textrm{-tagTi}}$  & Number of jets passing tight $c$-tag WP ($b$-veto) \\
& $N_{b\textrm{-tag70}}$  & Number of jets passing \SI{70}{\%} $b$-tag WP \\
& $N_{b\textrm{-tag60}}$  & Number of jets passing \SI{60}{\%} $b$-tag WP \\
\bottomrule
\end{tabular}
\label{Tab:BDTInputVariables}
\end{table}

The second category of variables involves event variables. The \NormProbLabel\ variable facilitates the identification of wrongly labelled jets
and can also reject non-\ttbar background. The \Njets\ and \ST\ variables are correlated with most top-quark kinematic variables.
For example, an event with small \bhad-jet \pt and large \ST\ suggests a signal event with a high-mass charged Higgs boson.

The third category contains flavour-tagging variables, which are the PCFT scores of the four labelled jets and the number
of jets passing a given PCFT working point. Of special interest are the PCFT scores of \ljet1 and \ljet2, as they
are sensitive to the different decay characteristics of the charged Higgs boson. While all signal events involve a $c$-quark in the boson decay,
only about \SI{50}{\%} of the \ttbar background events share the same characteristic. Other flavour-tagging variables are useful in rejecting
non-\ttbar background and add information via correlations with other variables.
The distributions of selected BDT input variables of different types and with large separation power are shown in Figure~\ref{Fig:BDTinputs}.

\begin{figure}
\subfloat[]{\includegraphics[width=0.49\textwidth]{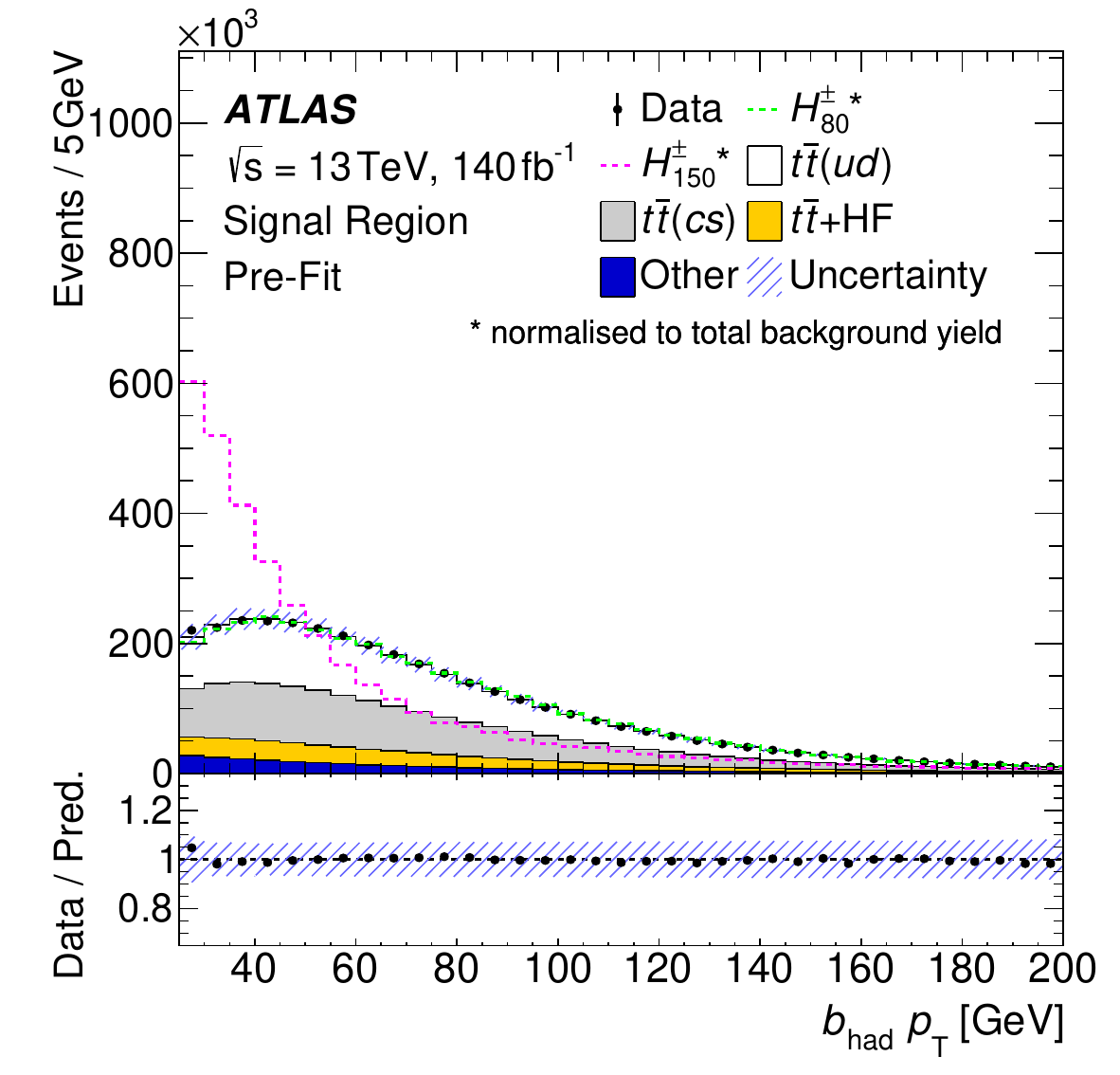}}
\subfloat[]{\includegraphics[width=0.49\textwidth]{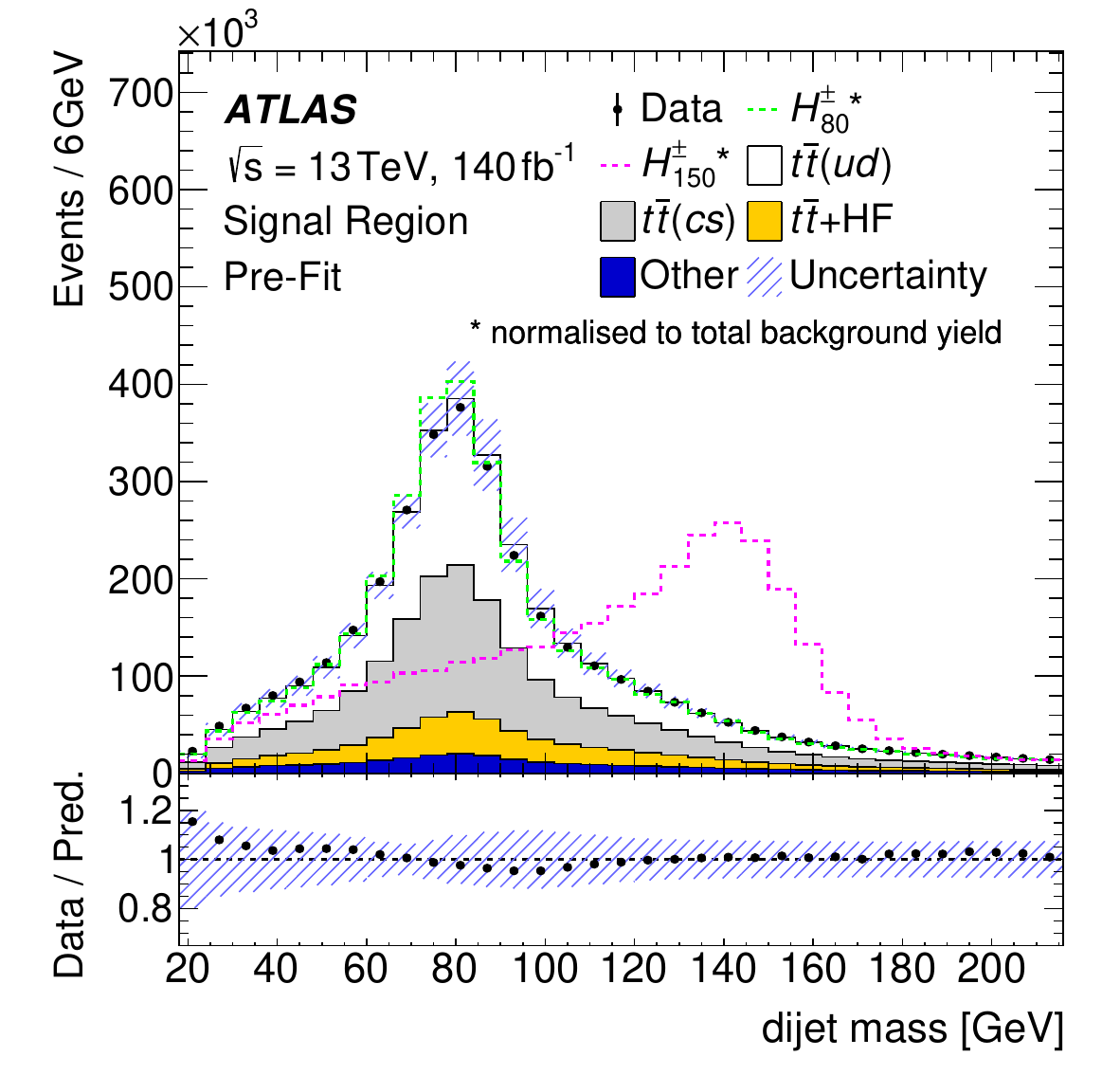}} \\
\subfloat[]{\includegraphics[width=0.49\textwidth]{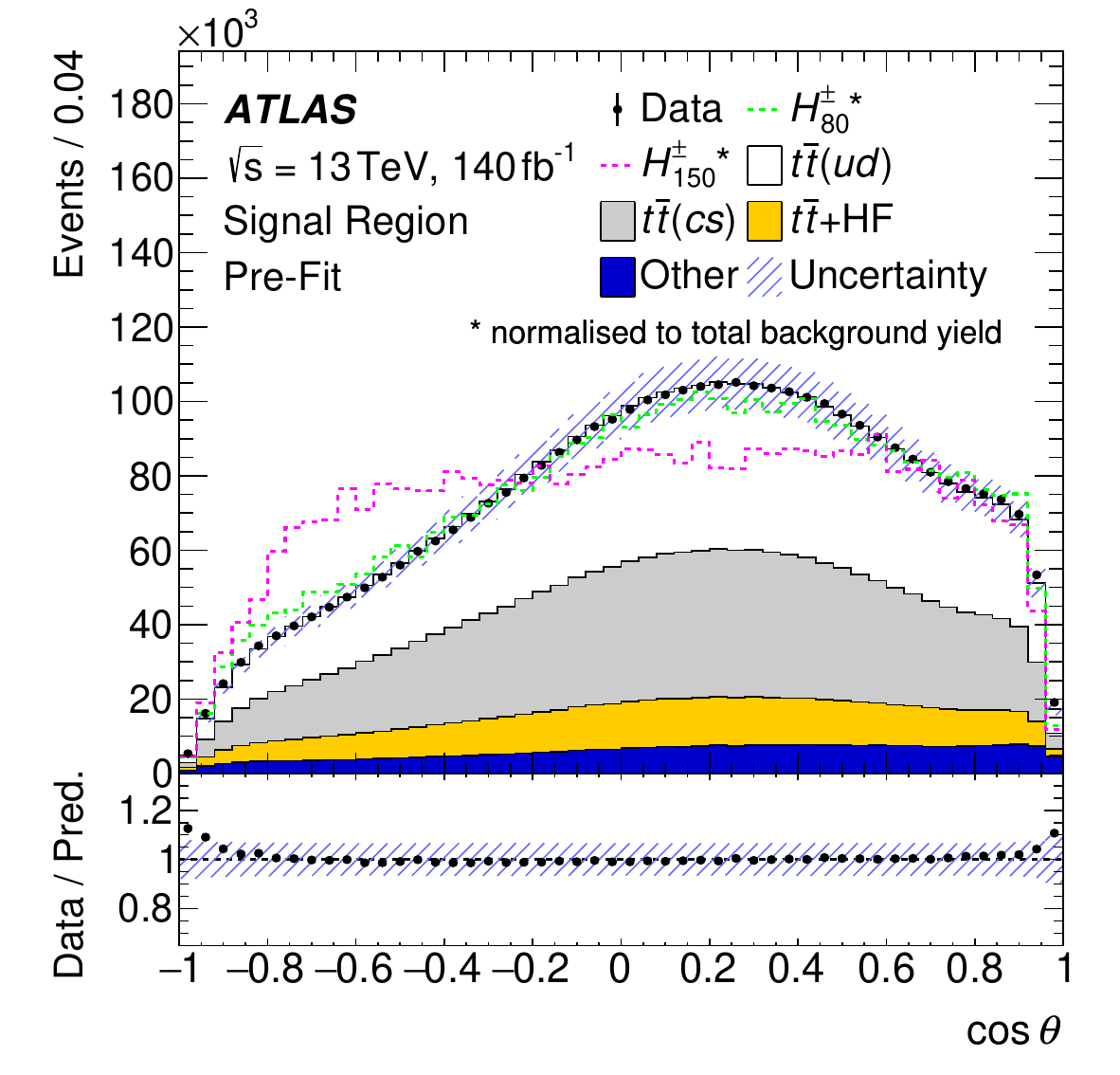}}
\subfloat[]{\includegraphics[width=0.49\textwidth]{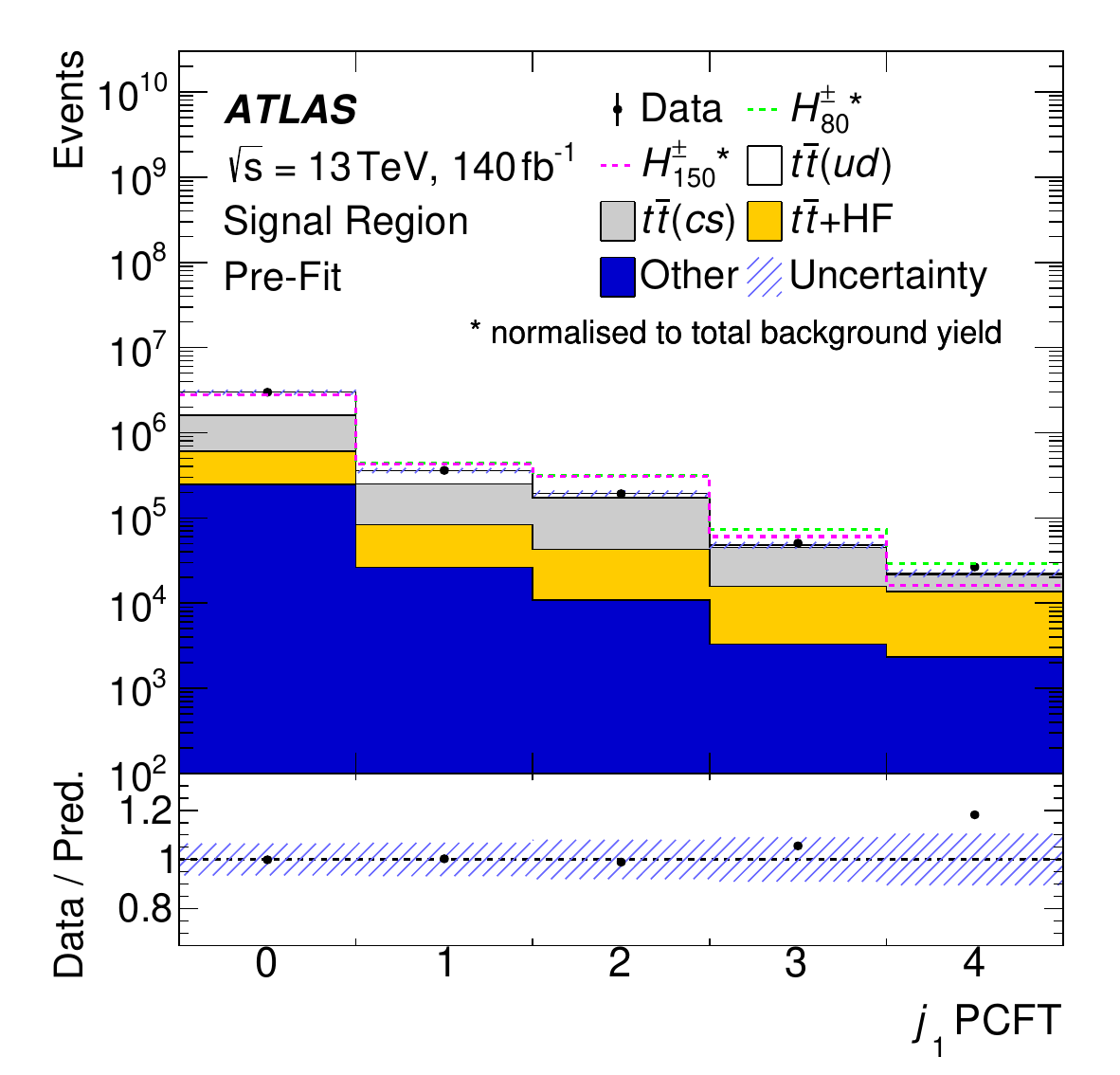}}
\caption{Distributions of selected BDT input variables. The variables are defined in Table~\ref{Tab:BDTInputVariables}.
The processes $\ttbar(\text{allHad})$, \singleTopWt, \singleTop, $\ttbar H$, \OtherTop, \Wjets, \Zjets, $VV$ listed in Table~\ref{Tab:Samples} are combined with the multijet background in the \enquote{Other} category.
The uncertainty band represents the combined statistical and systematic uncertainty of the prediction.
Overlaid are the shapes for the $\Hpm_{80}$ and $\Hpm_{150}$ signal samples normalised to the total background prediction.}
\label{Fig:BDTinputs}
\end{figure}

The optimisation of the BDT input variables was performed with a baseline set of hyperparameters -- including tree depth and learning rate
-- determined by a lightweight BDT hyperparameter scan. For the final training, the optimal configuration of the BDT hyperparameters
is determined for each signal mass point separately with the help of the \textit{hyperopt}~\cite{Bergstra_2015} tool. The \textit{hyperopt} tool
performs a Bayesian optimisation using Tree-structured Parzen Estimators to obtain the optimal parameter set from a given parameter range.
It is more efficient than grid or random searches because it uses previous training steps to learn where the optimum is going to be. To avoid any bias
between the hyperparameter optimisation dataset and the final test dataset, the scan was carried out using nested cross-training. For each of the
5-folds the scan is performed using 4-fold cross-training on the other four folds. For the final training the BDTs are retrained using all four folds as training data.
The BDT-score distributions after training with the $H^{\pm}_{80}$, $H^{\pm}_{110}$, $H^{\pm}_{130}$ and $H^{\pm}_{150}$ signal samples
are shown in Figure~\ref{Fig:BDTscores}.

\begin{figure}
\centering
\subfloat[]{\includegraphics[width=0.49\textwidth]{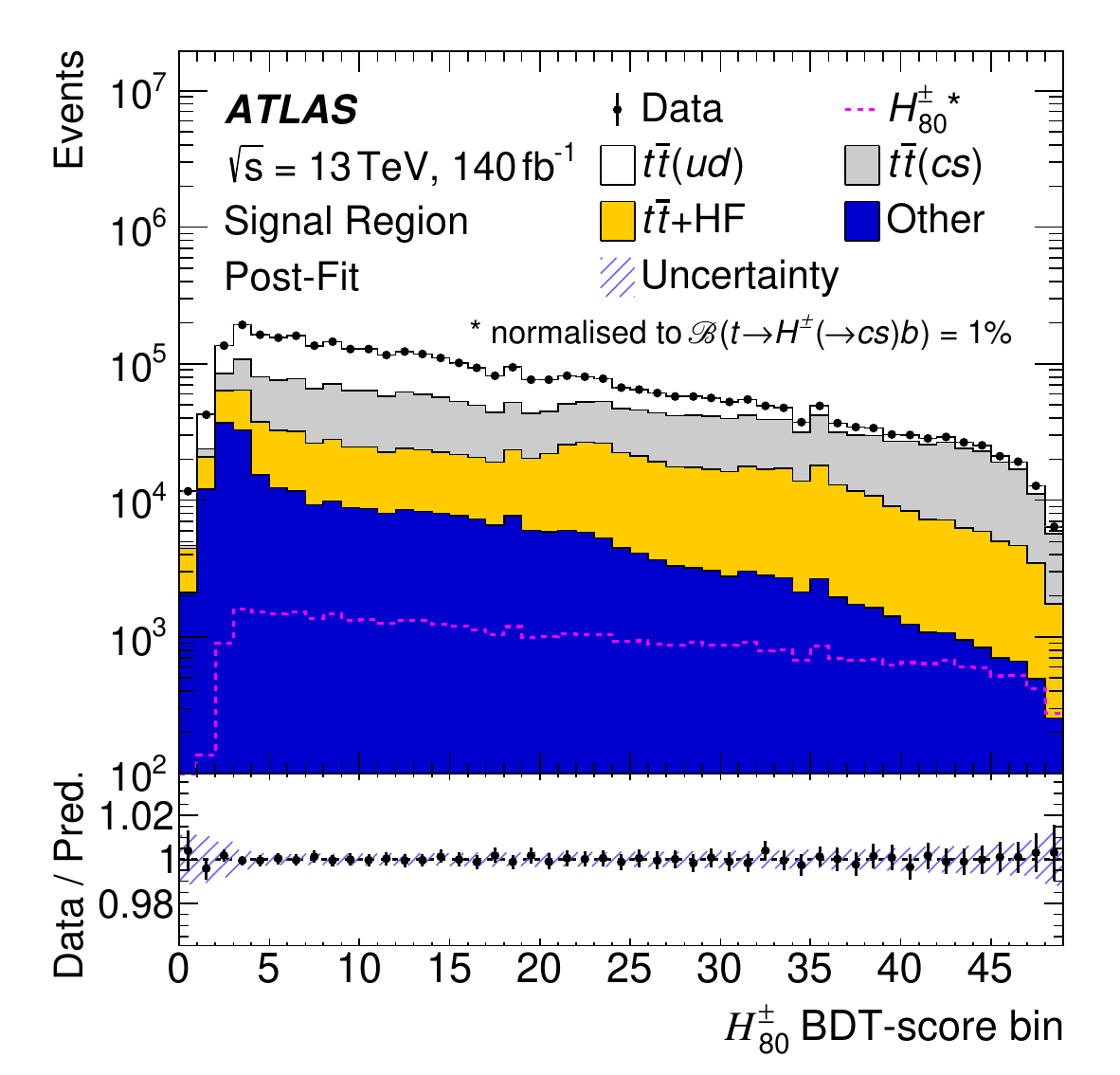}}
\subfloat[]{\includegraphics[width=0.49\textwidth]{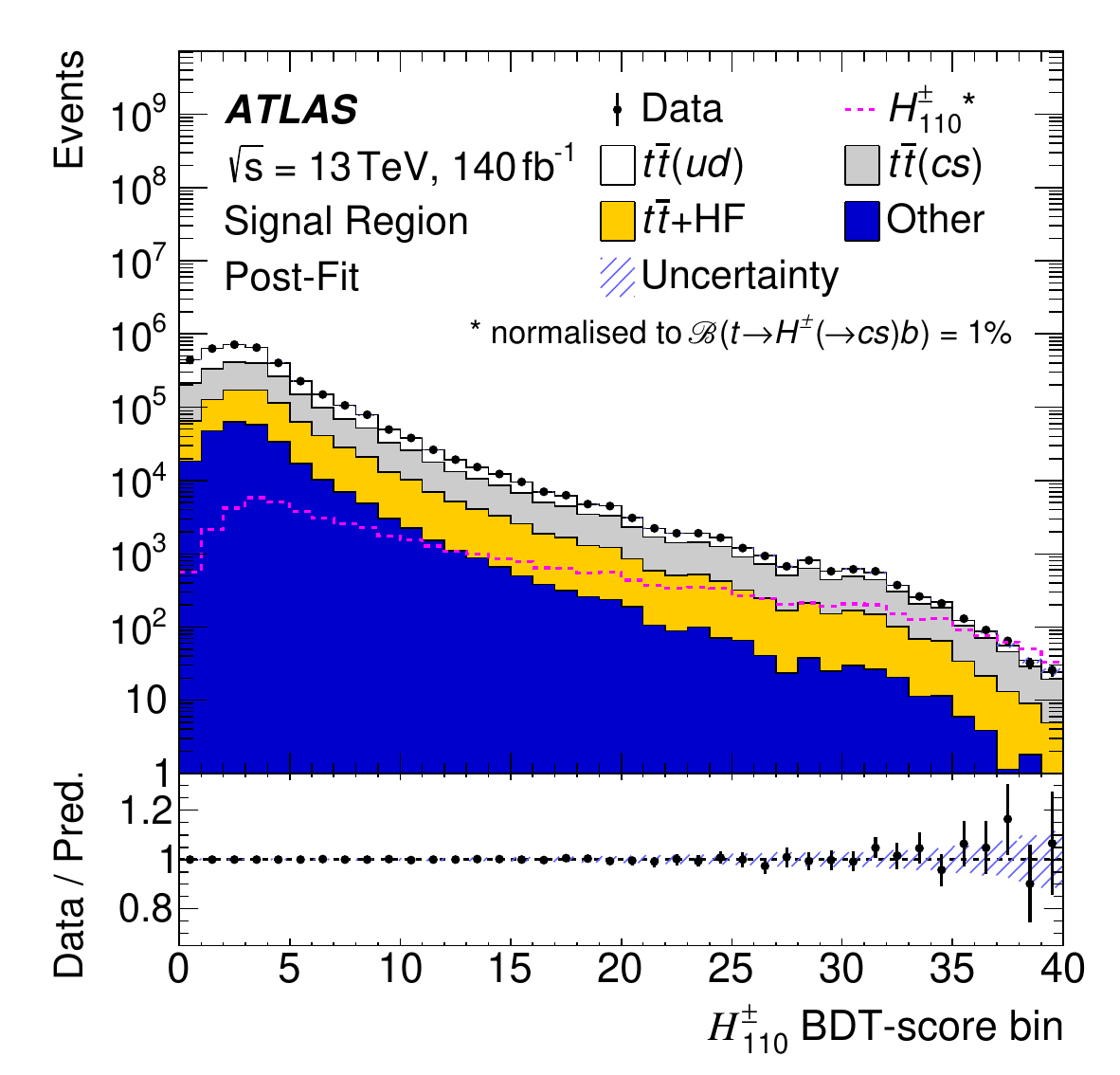}} \\
\subfloat[]{\includegraphics[width=0.49\textwidth]{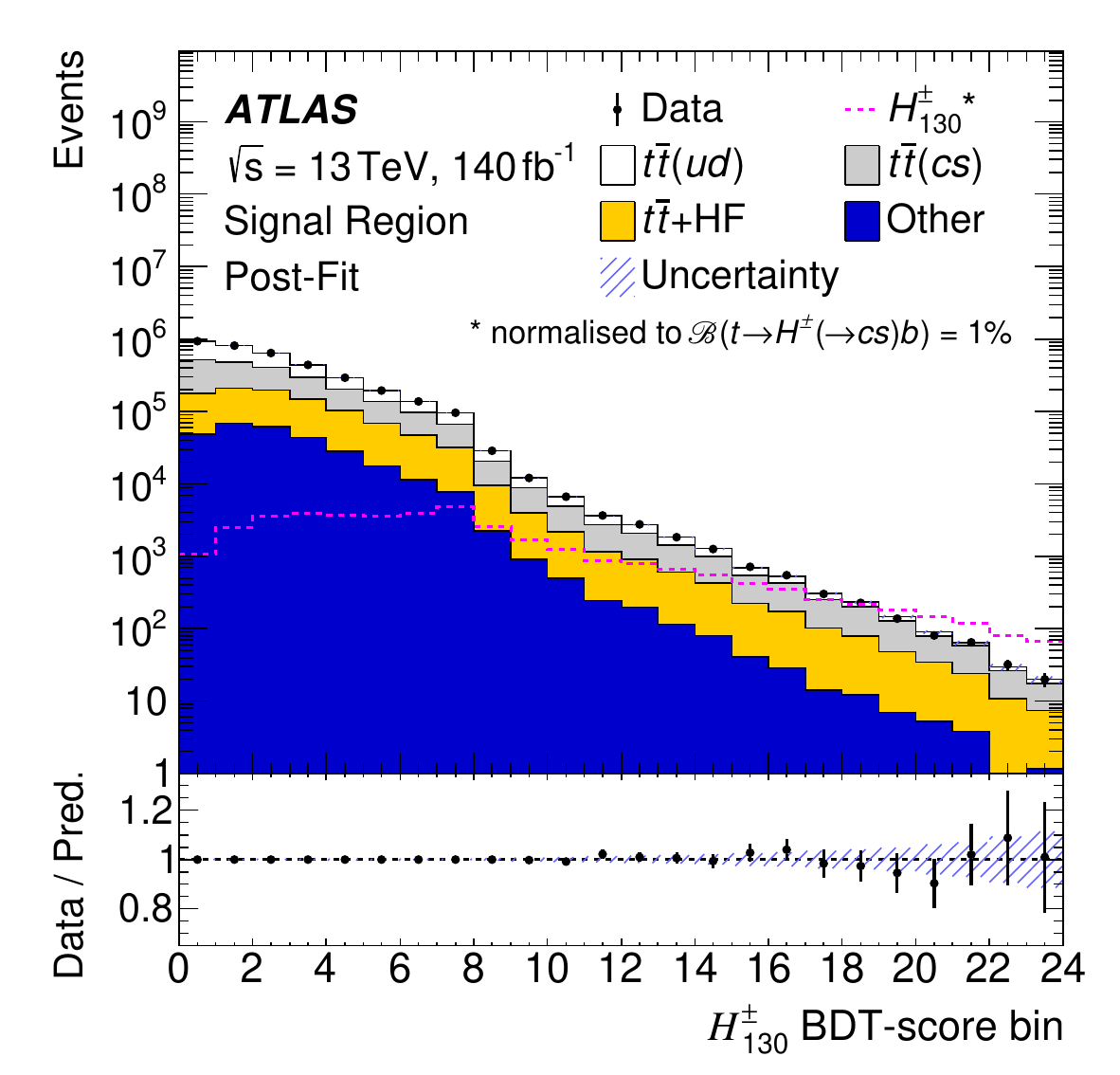}}
\subfloat[]{\includegraphics[width=0.49\textwidth]{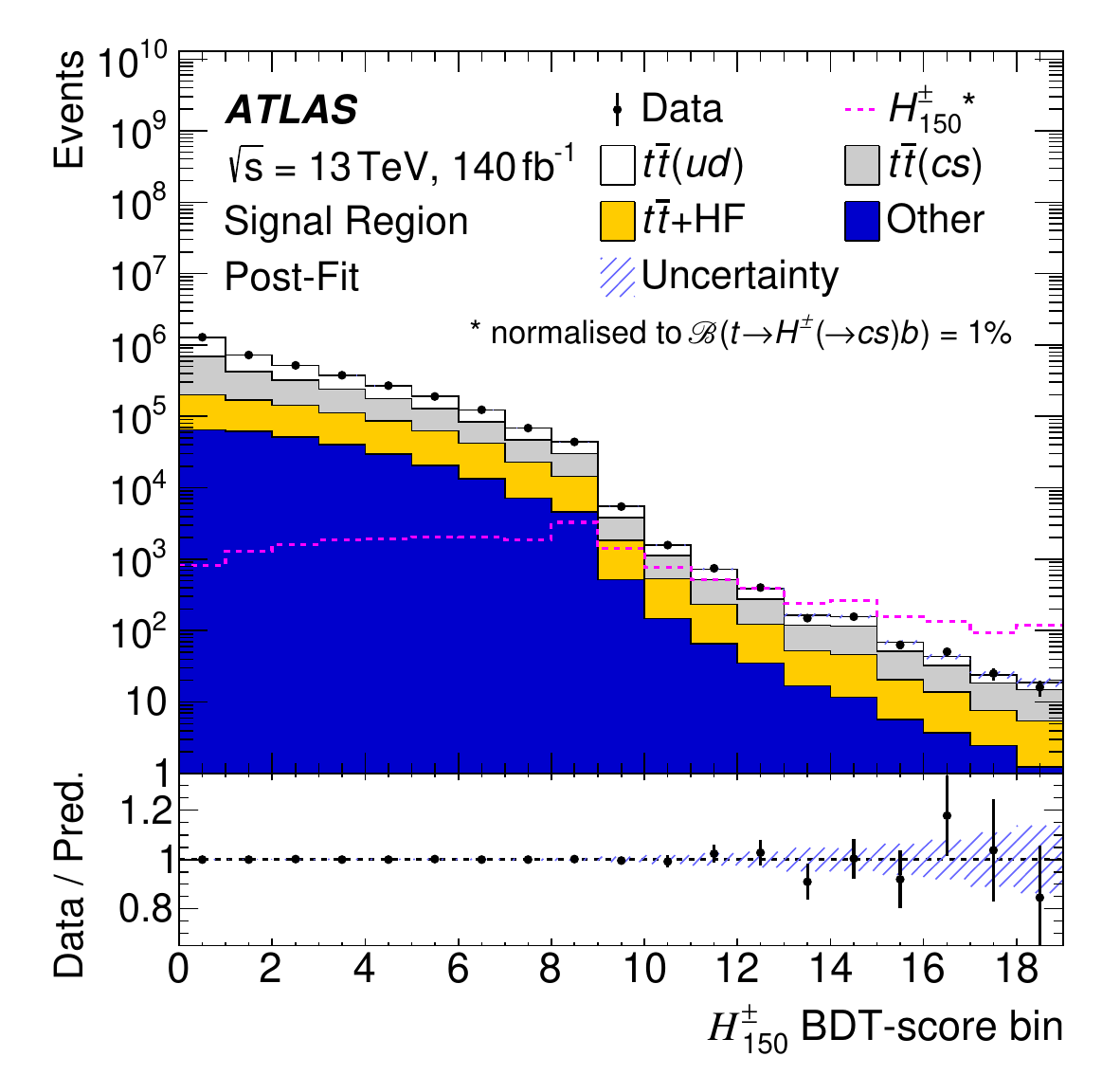}}
\caption{BDT-score distributions for the training with the (a) \SI{80}{\GeV}, (b) \SI{110}{\GeV}, (c) \SI{130}{\GeV} and (d) \SI{150}{\GeV}
signal mass hypotheses after performing background-only binned-likelihood fits to the distributions as described in Section~\ref{Sec:StatisticalInterpretation}.
The processes $\ttbar(\text{allHad})$, \singleTopWt, \singleTop, $\ttbar H$, \OtherTop, \Wjets, \Zjets, $VV$ listed in Table~\ref{Tab:Samples} are combined with the multijet background in the \enquote{Other} category.
The background yields are normalised to their best-fit values.
The uncertainty band represents the combined statistical and systematic uncertainty of the prediction.
Overlaid is the signal shape normalised to $\mathscr{B}(t \to \Hpm (\to cs) b) = \SI{1}{\%}$.
The binning procedure for the BDT score is defined in Section~\ref{Sec:StatisticalInterpretation}.}
\label{Fig:BDTscores}
\end{figure}


%
%
%

%
%
%

\section{Systematic uncertainties}
\label{Sec:Systematics}

This section discusses systematic uncertainties the analysis is sensitive to, including those affecting the detector response,
theoretical uncertainties, and modelling of signal and background processes that affect the normalisation and shapes of the
simulated signal and background distributions. The individual systematic uncertainties are considered to be uncorrelated,
while correlations for a given systematic uncertainty are maintained across signal and background processes.

The uncertainty in the combined 2015--2018 integrated luminosity is \SI{0.83}{\%}~\cite{DAPR-2021-01}, obtained using the
LUCID-2 detector~\cite{LUCID2} for the primary luminosity measurements, complemented by measurements using the inner detector and calorimeters.
This uncertainty is assigned to all physics processes whose normalisations are taken from simulation. An uncertainty in the correction of the \pileup
distribution~\cite{STDM-2015-05} in simulation to that in data is taken into account as well.

Uncertainties in the calibration of physics objects affect all simulated samples. Uncertainties associated with the trigger, reconstruction,
identification and isolation efficiency calibration~\cite{EGAM-2018-01,MUON-2018-03}, as well as the impact of
the energy (momentum) scale and resolution uncertainties~\cite{EGAM-2018-01,MUON-2022-01} on the selection efficiency are considered for
electrons (muons). Jet energy scale (JES)~\cite{PERF-2016-04} and resolution (JER)~\cite{PERF-2011-04} uncertainties and the uncertainty
in the efficiency of matching jets to the primary vertex~\cite{PERF-2014-03} are taken into account. The energy scale and resolution
uncertainties for leptons and jets are propagated to the \MET. In addition, the uncertainty in the \MET from tracks matched to the primary vertex but not to
other reconstructed objects~\cite{JETM-2020-03} is considered. Corrections to simulations to match the flavour-tagging and
mistagging efficiencies in data are taken into account in bins of the jet \pt~\cite{FTAG-2018-01,FTAG-2020-08,FTAG-2019-02}.

Uncertainties in the modelling of different processes are assumed to be uncorrelated. Modelling uncertainties for processes
making a small contribution to the total background yield ($\ttbar(\text{allHad})$, $\ttbar H$ and \OtherTop samples) are
neglected. Some modelling uncertainties are estimated in the same way for all processes. These are the uncertainties due to missing higher orders in the
perturbative expansion of the partonic cross-section, which are estimated by varying the renormalisation (\muR)
and factorisation (\muF) scales by a factor of two. The uncertainties associated with the choice of PDF are
evaluated by using dedicated PDF error eigensets. The PDF uncertainties are combined by calculating their standard deviation
with respect to the nominal set or by summing the differences in quadrature for Hessian sets. The uncertainty related to the
choice of strong coupling constant value \alphas is evaluated by comparing predictions using PDF sets obtained
with two alternative \alphas values. The uncertainties in the production cross-sections are included as normalisation
uncertainties~\cite{Martin:2009iq,Martin:2009bu,Aliev:2010zk,Kant:2014oha} for all processes whose normalisations are taken from simulation.

For top-quark processes the uncertainties associated with the choice of generator
for ME and PS simulation are assessed by comparing the nominal sample with alternative samples generated with \AMCatNLO
instead of \POWHEGBOX and with \HERWIG7 instead of \PYTHIA, respectively. The uncertainty due to the choice of
\hdamp parameter value is determined by comparison with an alternative sample generated with $\hdamp=3\,\mtop$ instead of $1.5\,\mtop$.
The alternative samples are described in detail in Section~\ref{Sec:DataMcSamples}.
The interference between $t \Wboson (b)$ and \ttbar processes~\cite{Frixione:2008yi} is handled using the diagram
removal scheme, which removes diagrams with intermediate top quarks. The uncertainty associated with this procedure is evaluated
with the help of an alternative \singleTopWt sample where a subtraction term is added to the matrix element to
cancel out the resonant top-quark pole contribution.

For top-quark and signal processes the uncertainties in the amounts of initial- and final-state QCD radiation (ISR and FSR) are estimated by
varying the corresponding parameter (Var3c) of the A14 PS tune and by varying the FSR scale ($\muR^\mathrm{FSR}$) by a factor of two, respectively.
The systematic uncertainty introduced by the \ttbar correction, described in Section~\ref{Sec:BackgroundModelling}, is estimated by performing an eigenvalue decomposition of the fitted parameters
and varying the eigenvalues separately by one standard deviation. Uncertainties related to the \ttbar ME, PS, \hdamp parameter and
FSR show, among the considered systematic uncertainties, the largest shape differences in the high BDT-score regions and
are therefore the dominant uncertainties in this search.

For weak-boson processes, electroweak corrections at next-to-leading order are estimated using the electroweak virtual approximation.
The electroweak and QCD components are combined using an exponentiated prescription~\cite{Kallweit:2014xda,PMGR-2021-01}.
CKKW and QSF are two parameters of \Sherpa that define the scale for merging/matching jets from the ME with the PS, and the
scale used for resummation of soft gluon emissions, respectively. Their impact on the BDT input observables is measured at
generator-level with the help of alternative samples in which the nominal values are varied. The observed differences are
covered by a \SI{17}{\%} normalisation uncertainty. The effect of NNLO correction factors on the cross-section
of single-boson processes is \SI{5}{\%}. This is added as a normalisation uncertainty to all weak-boson processes.

To evaluate the systematic uncertainty of the multijet background estimate, an alternative estimate is made in which,
instead of the lepton isolation criterion, a looser lepton identification is used to define the ABCD regions.
In addition, a conservative \SI{50}{\%} normalisation uncertainty is added to account for statistical
uncertainties of the transfer factor and MC uncertainties in regions \textit{B}, \textit{C} and \textit{D}.

The uncertainty in the expected event count in each bin, due to the finite MC sample sizes,
is accounted for by one Gaussian-constrained parameter per bin, which represents the total uncertainty of the MC event
content in that bin~\cite{Barlow:1993dm}.


%
%
%

%
%
%

\section{Statistical interpretation}
\label{Sec:StatisticalInterpretation}

The presence of a charged Higgs boson signal in data is quantified with the help of binned maximum-likelihood fits to the BDT-score distributions.
The statistical model is implemented using the Histfactory format~\cite{Cranmer:1456844}. Minimisation of the likelihood function
is performed in the pyhf framework~\cite{pyhf_joss}.

The parameter of interest is the branching fraction of the process $t \to H^\pm b$ (\BR), which is constrained to $\BR \in [0,0.1]$,\footnote{The upper bound on \BR is justified by measurements setting lower bounds on $\mathscr{B}(t \to Wb)$~\cite{D0:2010mwe}.}
while assuming $\mathscr{B}(t\to Wb) + \mathscr{B}(t \to \Hpm (\rightarrow cs)b)=1.0$. The \ttbar cross-section scaling factor (\mutt) and the fraction of
\ttbarHF events among \ttbar background events (\fHF) are unconstrained parameters in the fit, with nominal values of $1.0$ and
$0.1364$, respectively. These parameters relate the signal and background yields before the fit (pre-fit) and after the fit (post-fit) as follows:\footnote{
The event yields before the fit are set to the total expected $\ttbar$ yield, so that $N_{\Hpm}$(pre-fit) = $N_{\ttbarLF}$(pre-fit) = $N_{\ttbarHF}$(pre-fit).
For the nominal values of \mutt and \fHF, and $\BR=0$, one recovers the SM expectation.}
\begin{alignat*}{5}
&N_{\Hpm} (\text{post-fit})     &&= \mutt  &&                 &&\times 2 (1-\BR) \,\, \BR &&\times N_{\Hpm} (\text{pre-fit}), \\
&N_{\ttbarLF} (\text{post-fit}) &&= \mutt  &&\times  (1-\fHF) &&\times (1-\BR)^{2}             &&\times N_{\ttbarLF} (\text{pre-fit}), \\
&N_{\ttbarHF} (\text{post-fit}) &&= \mutt  &&\times  \fHF     &&\times (1-\BR)^{2}             &&\times N_{\ttbarHF} (\text{pre-fit}).
\end{alignat*}

The systematic uncertainties described in Section~\ref{Sec:Systematics} are implemented as Gaussian-constrained
nuisance parameters (NPs) in the fit. Systematic uncertainties with just one component are symmetrised by mirroring the nominal template.
All two-point systematic uncertainties and some reweighting systematic uncertainties with large statistical fluctuations are smoothed.
Uncertainties with a negligible impact on the uncertainty of \BR are removed from the likelihood fit to improve numerical performance.

The discovery significance of a signal in data is calculated in a likelihood ratio test where the background-only hypothesis is compared
with the signal-plus-background hypothesis~\cite{Cousins:2007bmb}. The asymptotic approximation is used to
estimate the probability distribution of the test statistic~\cite{Cowan:2010js}. Upper limits on \BR are set by
determining the \BR value which can be rejected in \SI{95}{\%} of the cases, i.e.\ at \SI{95}{\%} confidence level (CL), with respect to the
best-fit signal-plus-background hypothesis. The modified frequentist technique (CL$_\text{s}$)~\cite{Read:2002hq} is used to avoid
excluding signal models where the analysis has little sensitivity. The median upper limit, referred to as the expected upper limit, and its
$1\sigma$ and $2\sigma$ expected variations are derived from a background-only Asimov dataset~\cite{Cowan:2010js}.

The BDT-score templates are binned in such a way that the number of bins is minimised, while maximising the expected sensitivity and ensuring
the stability of the fit and the validity of the asymptotic approximation. This is achieved by starting from templates with 10\,000 equal-width
bins and iteratively merging bins from right to left -- starting from the most discriminating bins -- until a bin in question fulfils all conditions.
Each bin is required to contain at least 20 expected background events, and the signal and background MC statistical uncertainties have to be
less than \SI{20}{\%} and \SI{10}{\%}, respectively. Each bin has to pass a given $S/\sqrt{B}$ threshold defined by the
$S/\sqrt{B}$ ratio\footnote{$S$ is the signal and $B$ the total background prediction in the respective bin.} in the first merged bin.
In order to retain shape information in the low discriminating BDT-score region, a maximum of 1000 bins may be merged into a single bin.
Additionally, the $S/\sqrt{B}$ threshold may be adjusted for BDT scores with very high or very low separation power such that each template has at least 19 and no more than 49 bins.


%
%
%
%
%
%

%
%
%

\section{Results}
\label{Sec:Results}

The binned maximum-likelihood fit described in Section~\ref{Sec:StatisticalInterpretation} is performed per signal mass-point hypothesis
on the respective BDT-score distribution. Table~\ref{Tab:Yields} shows the data and background yields after the background-only fit
of the BDT-score distribution trained with the \SI{130}{\GeV} signal mass point.
\begin{table}
\sisetup{group-minimum-digits=3,round-mode=figures,round-precision=2}
\centering
\caption{Data and background yields after the background-only fit of the BDT-score distribution for the \SI{130}{\GeV} signal mass BDT training.
For comparison, the expected signal yield for $\BR=\SI{1}{\%}$ is added.
The sample names are defined in Table~\ref{Tab:Samples}.}
\begin{tabular}{l r@{\(\;\pm\;\)}r r@{\(\;\pm\;\)}r}
\toprule
Name & \multicolumn{2}{c}{Post-fit yields} \\
\midrule
\ttbarll      & \num{1366035} & \num{ 76355} \\
\ttbarlc      & \num{1247183} & \num{ 92165} \\
\ttbarHF      & \num{ 711901} & \num{146110} \\
\singleTopWt  & \num{ 104962} & \num{ 22671} \\
\singleTop    & \num{  68059} & \num{ 28237} \\
\Wjets        & \num{  70348} & \num{ 28671} \\
\Zjets \& $VV$& \num{  20940} & \num{  9480} \\
\OtherTop \& $\ttbar(\text{allHad})$ \& $\ttbar H$  & \num{  16779} & \num{   452} \\
Multijet (MJ) & \num{  11653} & \num{  6805} \\
\hline
Total background & \num{3618340} & \num{ 10908} \\
Data          & \multicolumn{2}{l}{\num{3618340}} \\
\hline
$\Hpm_{130}$ ($\BR=\SI{1}{\%}$)  & \multicolumn{2}{l}{$\,\,\,\,\,\,\,$\num{37986}}\\
\bottomrule
\end{tabular}
\label{Tab:Yields}
\end{table}

The fitted \mutt values agree within their uncertainties with the SM prediction of $1.0$ for all mass points.
The \fHF parameter is measured to be \num{0.19 \pm 0.02} for the $\Hpm_{80}$ fit, which is the most precise fit.
The measured values of \fHF in other fits agree with this value within their uncertainties.
This is a larger heavy-flavour fraction than predicted by simulations, but agrees with many other ATLAS analyses~\cite{TOPQ-2014-10,TOPQ-2017-12,CMS-TOP-18-002,HDBS-2018-51,HDBS-2019-24,TOPQ-2020-10}.
The fitted \BR values are equal to, or compatible with, zero for most signal mass hypotheses.
The largest signal significance observed in data is for the \SI{110}{\GeV} mass point, with a local $p$-value of \SI{5}{\%} ($1.5\sigma$).

The impact of systematic uncertainties on the \BR measurement's accuracy is estimated by fixing the NP under consideration to its post-fit value,
performing the fit and comparing the uncertainty of the fitted \BR with the one from the nominal fit. The results of these fits when
fixing a group of systematic uncertainties are summarised in Table~\ref{Tab:SystematicsImpact}. For the \SI{80}{\GeV} signal hypothesis the kinematics
are very similar for the signal and \ttbar processes. Therefore, the flavour-tagging uncertainties have the largest impact on \BR. For other signal hypotheses the
\ttbar modelling NPs, especially those for the ME, PS, FSR and \hdamp uncertainties, have the largest impact on the \BR uncertainty.
MC statistical uncertainties naturally become more important with finer BDT-score binning.
Less impactful but still important are jet, single-top-quark and weak-boson modelling uncertainties. Least impactful are luminosity, \pileup, lepton and \MET-related uncertainties.
\begin{table}
\centering
\caption{Breakdown of the relative contributions to the uncertainty in the extracted \BR in the likelihood fit to data.
The contributions are obtained by fixing the relevant NPs to their post-fit values in the likelihood fit.
The square root of the difference of the squares of the nominal uncertainty and obtained uncertainty is divided by the nominal uncertainty to obtain the relative impact.
The sum in quadrature of the individual components differs from the total uncertainty due to correlations between uncertainties in the different groups.
The uncertainty from data statistical uncertainties is determined from fits with all NPs fixed to their post-fit values.
The total uncertainty in \BR for the fits with $\Hpm_{80}$ and $\Hpm_{150}$ is \SI{1.2}{\%} and \SI{0.04}{\%}, respectively.}
\resizebox{\textwidth}{!}{
\begin{tabular}{l S | l S}
\toprule
\multicolumn{2}{c|}{$\Hpm_{80}$} & \multicolumn{2}{c}{$\Hpm_{150}$}\\
Category                        & {Relative contribution} & Category                            & {Relative contribution} \\
\midrule
\textbf{Data statistical}           & ~~\SI{6}{\%}    & \textbf{Data statistical}           & \SI{38}{\%} \\
\textbf{Systematic}                 & ~~~\SI{99.8}{\%} & \textbf{Systematic}                 & \SI{93}{\%} \\
$\quad$ Flavour-tagging             & \SI{64}{\%}   & $\quad$ \ttbar modelling            & \SI{72}{\%} \\
$\quad$ MC statistical              & \SI{64}{\%}   & $\quad$ MC statistical              & \SI{35}{\%} \\
$\quad$ \ttbar modelling            & \SI{50}{\%}   & $\quad$ Weak-boson \& MJ modelling  & \SI{27}{\%} \\
$\quad$ \mutt \& \fLF               & \SI{21}{\%}   & $\quad$ Single-top-quark modelling  & \SI{25}{\%} \\
$\quad$ Jet                         & \SI{19}{\%}   & $\quad$ \mutt \& \fLF               & \SI{24}{\%} \\
$\quad$ Single-top-quark modelling  & \SI{16}{\%}   & $\quad$ Jet                         & \SI{23}{\%} \\
$\quad$ Luminosity \& \pileup       & \SI{15}{\%}   & $\quad$ Flavour-tagging             & \SI{20}{\%} \\
$\quad$ Weak-boson \& MJ modelling  & \SI{12}{\%}   & $\quad$ Lepton \& \MET              & ~~\SI{8}{\%}  \\
$\quad$ Signal modelling            & ~~\SI{8}{\%}    & $\quad$ Luminosity \& \pileup       & ~~\SI{7}{\%}  \\
$\quad$ Lepton \& \MET              & ~~\SI{7}{\%}    & $\quad$ Signal modelling            & ~~\SI{5}{\%}  \\
\bottomrule
\end{tabular}
}
\label{Tab:SystematicsImpact}
\end{table}

Figure~\ref{Fig:Limits} shows the expected limits on \BR with their $1\sigma$ and $2\sigma$
uncertainty bands. The expected limits are least stringent for the \SI{80}{\GeV} mass point, at about \SI{2.3}{\%}, as this signal mass point is closest to
the \Wboson-boson mass. For this mass point, the flavour-tagging information is the most powerful discriminant between signal and background.
Moving away from the \Wboson-boson mass the top-quark kinematics start to differ more and the limits improve.
The most stringent limits are expected for the \SI{150}{\GeV} mass point, at about \SI{0.077}{\%}. Close to the top-quark mass
threshold the limits weaken again as the acceptance decreases. The small acceptance is caused by the low average momentum of the \bhad-quark so that the resulting jets often fail the kinematic
requirements. The solid line in Figure~\ref{Fig:Limits} represents the observed limits.
The observed limits vary between \SI{0.066}{\%} to \SI{3.6}{\%}. Expected and observed limits agree within uncertainties.

\begin{figure}
\centering
\includegraphics[width=0.7\textwidth]{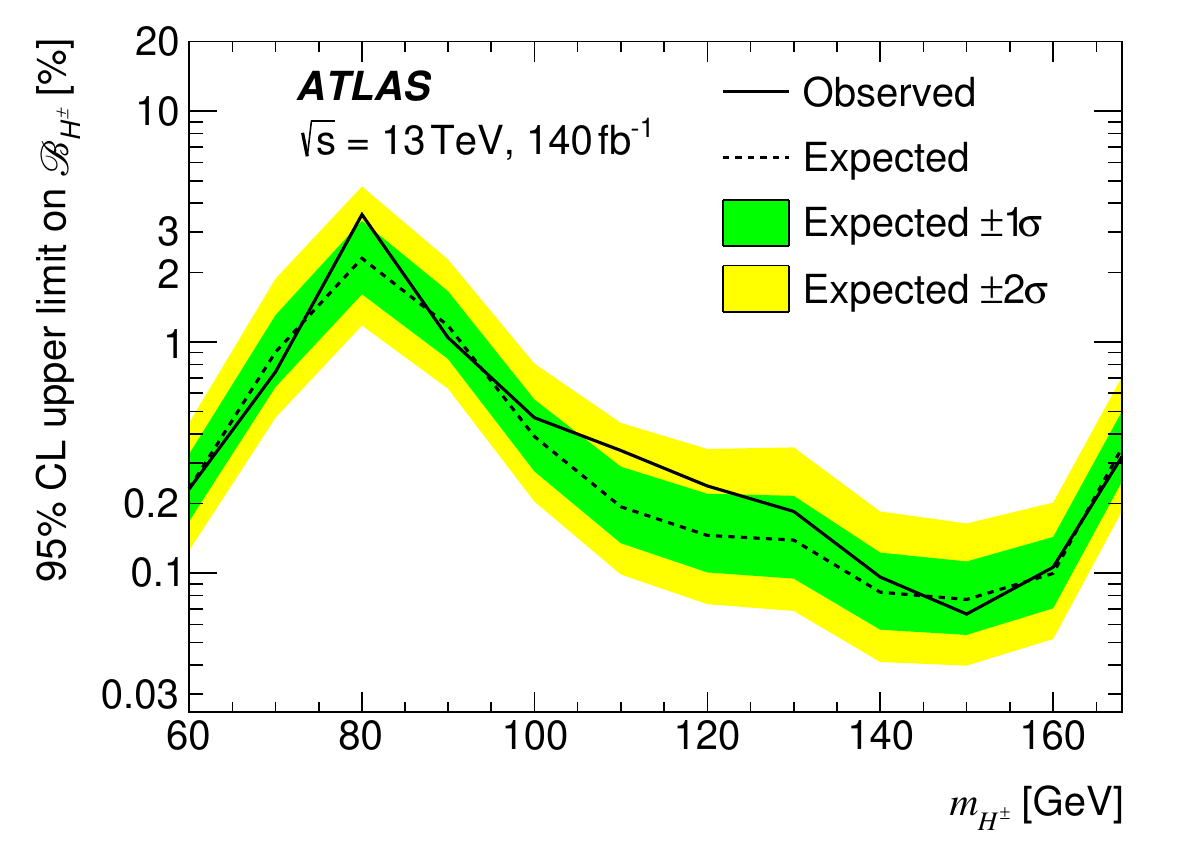}
\caption{Observed (solid line) and expected (dotted line) upper limits on \BR for charged Higgs boson with masses between \SI{60}{\GeV} and \SI{168}{\GeV}, assuming $\mathscr{B}(t\to Wb) + \mathscr{B}(t \to \Hpm (\rightarrow cs)b)=1.0$.
The $\pm 1 \sigma$ and $\pm 2 \sigma$ variations around the expected upper limit are indicated by the green and yellow bands, respectively.}
\label{Fig:Limits}
\end{figure}

%
%
%
%

%
%
%
%
%
%
%
%


%
%
\FloatBarrier

%
%
%

\section{Conclusions}
\label{Sec:Conclusion}

A search for a light charged Higgs boson produced in decays of the top quark, $t \to \Hpm b$, with $\Hpm \to cs$
is performed in the \Hpm mass range from 60 to \SI{168}{\GeV}. The data analysed corresponds to $\SI{140}{\ifb}$
of \pp collisions at $\sqrt{s} = \SI{13}{\TeV}$ recorded with the ATLAS detector at the LHC between 2015 and 2018.
This analysis focuses on the lepton-plus-jets final state, characterised by an isolated electron or muon and at
least four jets. The search exploits $b$-quark and $c$-quark identification techniques as well as multivariate methods
to suppress the dominant \ttbar background. No significant signal excess is found in data. Observed and expected \SI{95}{\%}
CL upper limits on the branching fraction $\mathscr{B}(t\rightarrow \Hpm b)$, assuming
$\mathscr{B}(t\to Wb) + \mathscr{B}(t \to \Hpm (\rightarrow cs)b)=1.0$, are found to range from \SI{0.066}{\%} to \SI{3.6}{\%} and \SI{0.077}{\%} to \SI{2.3}{\%},
respectively, depending on the mass of the charged Higgs boson.
These are the first direct limits on $\mathscr{B}(t\rightarrow \Hpm b)$ in the $\Hpm \rightarrow cs$ channel for charged Higgs bosons with masses
of \SI{60}{\GeV}, \SI{70}{\GeV} and \SI{168}{\GeV}, and currently the most stringent limits for masses between \SI{120}{\GeV} and \SI{160}{\GeV}.

%
%
%


%

%
\section*{Acknowledgements}
%

%
%

%
%

We thank CERN for the very successful operation of the LHC and its injectors, as well as the support staff at
CERN and at our institutions worldwide without whom ATLAS could not be operated efficiently.

The crucial computing support from all WLCG partners is acknowledged gratefully, in particular from CERN, the ATLAS Tier-1 facilities at TRIUMF/SFU (Canada), NDGF (Denmark, Norway, Sweden), CC-IN2P3 (France), KIT/GridKA (Germany), INFN-CNAF (Italy), NL-T1 (Netherlands), PIC (Spain), RAL (UK) and BNL (USA), the Tier-2 facilities worldwide and large non-WLCG resource providers. Major contributors of computing resources are listed in Ref.~\cite{ATL-SOFT-PUB-2023-001}.

We gratefully acknowledge the support of ANPCyT, Argentina; YerPhI, Armenia; ARC, Australia; BMWFW and FWF, Austria; ANAS, Azerbaijan; CNPq and FAPESP, Brazil; NSERC, NRC and CFI, Canada; CERN; ANID, Chile; CAS, MOST and NSFC, China; Minciencias, Colombia; MEYS CR, Czech Republic; DNRF and DNSRC, Denmark; IN2P3-CNRS and CEA-DRF/IRFU, France; SRNSFG, Georgia; BMBF, HGF and MPG, Germany; GSRI, Greece; RGC and Hong Kong SAR, China; ISF and Benoziyo Center, Israel; INFN, Italy; MEXT and JSPS, Japan; CNRST, Morocco; NWO, Netherlands; RCN, Norway; MNiSW, Poland; FCT, Portugal; MNE/IFA, Romania; MSTDI, Serbia; MSSR, Slovakia; ARIS and MVZI, Slovenia; DSI/NRF, South Africa; MICIU/AEI, Spain; SRC and Wallenberg Foundation, Sweden; SERI, SNSF and Cantons of Bern and Geneva, Switzerland; NSTC, Taipei; TENMAK, T\"urkiye; STFC/UKRI, United Kingdom; DOE and NSF, United States of America.

Individual groups and members have received support from BCKDF, CANARIE, CRC and DRAC, Canada; CERN-CZ, FORTE and PRIMUS, Czech Republic; COST, ERC, ERDF, Horizon 2020, ICSC-NextGenerationEU and Marie Sk{\l}odowska-Curie Actions, European Union; Investissements d'Avenir Labex, Investissements d'Avenir Idex and ANR, France; DFG and AvH Foundation, Germany; Herakleitos, Thales and Aristeia programmes co-financed by EU-ESF and the Greek NSRF, Greece; BSF-NSF and MINERVA, Israel; NCN and NAWA, Poland; La Caixa Banking Foundation, CERCA Programme Generalitat de Catalunya and PROMETEO and GenT Programmes Generalitat Valenciana, Spain; G\"{o}ran Gustafssons Stiftelse, Sweden; The Royal Society and Leverhulme Trust, United Kingdom.

In addition, individual members wish to acknowledge support from Armenia: Yerevan Physics Institute (FAPERJ); CERN: European Organization for Nuclear Research (CERN PJAS); Chile: Agencia Nacional de Investigaci\'on y Desarrollo (FONDECYT 1230812, FONDECYT 1230987, FONDECYT 1240864); China: Chinese Ministry of Science and Technology (MOST-2023YFA1605700), National Natural Science Foundation of China (NSFC - 12175119, NSFC 12275265, NSFC-12075060); Czech Republic: Czech Science Foundation (GACR - 24-11373S), Ministry of Education Youth and Sports (FORTE CZ.02.01.01/00/22\_008/0004632), PRIMUS Research Programme (PRIMUS/21/SCI/017); EU: H2020 European Research Council (ERC - 101002463); European Union: European Research Council (ERC - 948254, ERC 101089007), Horizon 2020 Framework Programme (MUCCA - CHIST-ERA-19-XAI-00), European Union, Future Artificial Intelligence Research (FAIR-NextGenerationEU PE00000013), Italian Center for High Performance Computing, Big Data and Quantum Computing (ICSC, NextGenerationEU); France: Agence Nationale de la Recherche (ANR-20-CE31-0013, ANR-21-CE31-0013, ANR-21-CE31-0022, ANR-22-EDIR-0002), Investissements d'Avenir Labex (ANR-11-LABX-0012); Germany: Baden-Württemberg Stiftung (BW Stiftung-Postdoc Eliteprogramme), Deutsche Forschungsgemeinschaft (DFG - 469666862, DFG - CR 312/5-2); Italy: Istituto Nazionale di Fisica Nucleare (ICSC, NextGenerationEU), Ministero dell'Università e della Ricerca (PRIN - 20223N7F8K - PNRR M4.C2.1.1); Japan: Japan Society for the Promotion of Science (JSPS KAKENHI JP22H01227, JSPS KAKENHI JP22H04944, JSPS KAKENHI JP22KK0227, JSPS KAKENHI JP23KK0245); Netherlands: Netherlands Organisation for Scientific Research (NWO Veni 2020 - VI.Veni.202.179); Norway: Research Council of Norway (RCN-314472); Poland: Ministry of Science and Higher Education (IDUB AGH, POB8, D4 no 9722), Polish National Agency for Academic Exchange (PPN/PPO/2020/1/00002/U/00001), Polish National Science Centre (NCN 2021/42/E/ST2/00350, NCN OPUS nr 2022/47/B/ST2/03059, NCN UMO-2019/34/E/ST2/00393, NCN \& H2020 MSCA 945339, UMO-2020/37/B/ST2/01043, UMO-2021/40/C/ST2/00187, UMO-2022/47/O/ST2/00148, UMO-2023/49/B/ST2/04085, UMO-2023/51/B/ST2/00920); Slovenia: Slovenian Research Agency (ARIS grant J1-3010); Spain: Generalitat Valenciana (Artemisa, FEDER, IDIFEDER/2018/048), Ministry of Science and Innovation (MCIN \& NextGenEU PCI2022-135018-2, MICIN \& FEDER PID2021-125273NB, RYC2019-028510-I, RYC2020-030254-I, RYC2021-031273-I, RYC2022-038164-I), PROMETEO and GenT Programmes Generalitat Valenciana (CIDEGENT/2019/027); Sweden: Swedish Research Council (Swedish Research Council 2023-04654, VR 2018-00482, VR 2022-03845, VR 2022-04683, VR 2023-03403, VR grant 2021-03651), Knut and Alice Wallenberg Foundation (KAW 2018.0157, KAW 2018.0458, KAW 2019.0447, KAW 2022.0358); Switzerland: Swiss National Science Foundation (SNSF - PCEFP2\_194658); United Kingdom: Leverhulme Trust (Leverhulme Trust RPG-2020-004), Royal Society (NIF-R1-231091); United States of America: U.S. Department of Energy (ECA DE-AC02-76SF00515), Neubauer Family Foundation.

%
%


%
%
%
%
%
%
%

%
%
%
%
%
%

%
%

%
%
%
\printbibliography
\clearpage
 
\begin{flushleft}
\hypersetup{urlcolor=black}
{\Large The ATLAS Collaboration}

\bigskip

\AtlasOrcid[0000-0002-6665-4934]{G.~Aad}$^\textrm{\scriptsize 104}$,
\AtlasOrcid[0000-0001-7616-1554]{E.~Aakvaag}$^\textrm{\scriptsize 17}$,
\AtlasOrcid[0000-0002-5888-2734]{B.~Abbott}$^\textrm{\scriptsize 123}$,
\AtlasOrcid[0000-0002-0287-5869]{S.~Abdelhameed}$^\textrm{\scriptsize 119a}$,
\AtlasOrcid[0000-0002-1002-1652]{K.~Abeling}$^\textrm{\scriptsize 56}$,
\AtlasOrcid[0000-0001-5763-2760]{N.J.~Abicht}$^\textrm{\scriptsize 50}$,
\AtlasOrcid[0000-0002-8496-9294]{S.H.~Abidi}$^\textrm{\scriptsize 30}$,
\AtlasOrcid[0009-0003-6578-220X]{M.~Aboelela}$^\textrm{\scriptsize 45}$,
\AtlasOrcid[0000-0002-9987-2292]{A.~Aboulhorma}$^\textrm{\scriptsize 36e}$,
\AtlasOrcid[0000-0001-5329-6640]{H.~Abramowicz}$^\textrm{\scriptsize 155}$
% ...
\bigskip
\\

% ...
$^{17}$Department for Physics and Technology, University of Bergen, Bergen; Norway.\\
% ...
$^{30}$Physics Department, Brookhaven National Laboratory, Upton NY; United States of America.\\
% ...
$^{36}$$^{(a)}$Facult\'e des Sciences Ain Chock, Universit\'e Hassan II de Casablanca;$^{(b)}$Facult\'{e} des Sciences, Universit\'{e} Ibn-Tofail, K\'{e}nitra;$^{(c)}$Facult\'e des Sciences Semlalia, Universit\'e Cadi Ayyad, LPHEA-Marrakech;$^{(d)}$LPMR, Facult\'e des Sciences, Universit\'e Mohamed Premier, Oujda;$^{(e)}$Facult\'e des sciences, Universit\'e Mohammed V, Rabat;$^{(f)}$Institute of Applied Physics, Mohammed VI Polytechnic University, Ben Guerir; Morocco.\\
% ...
$^{45}$Physics Department, Southern Methodist University, Dallas TX; United States of America.\\
% ...
$^{50}$Fakult\"{a}t Physik , Technische Universit{\"a}t Dortmund, Dortmund; Germany.\\
% ...
$^{56}$II. Physikalisches Institut, Georg-August-Universit\"{a}t G\"ottingen, G\"ottingen; Germany.\\
% ...
$^{104}$CPPM, Aix-Marseille Universit\'e, CNRS/IN2P3, Marseille; France.\\
% ...
$^{119}$$^{(a)}$New York University Abu Dhabi, Abu Dhabi;$^{(b)}$United Arab Emirates University, Al Ain; United Arab Emirates.\\
% ...
$^{123}$Homer L. Dodge Department of Physics and Astronomy, University of Oklahoma, Norman OK; United States of America.\\
% ...
$^{155}$Raymond and Beverly Sackler School of Physics and Astronomy, Tel Aviv University, Tel Aviv; Israel.\\
% ...
$^{a}$ Also Affiliated with an institute covered by a cooperation agreement with CERN.\\
$^{b}$ Also at An-Najah National University, Nablus; Palestine.\\
$^{c}$ Also at Borough of Manhattan Community College, City University of New York, New York NY; United States of America.\\
$^{d}$ Also at Center for Interdisciplinary Research and Innovation (CIRI-AUTH), Thessaloniki; Greece.\\
$^{e}$ Also at CERN, Geneva; Switzerland.\\
$^{f}$ Also at CMD-AC UNEC Research Center, Azerbaijan State University of Economics (UNEC); Azerbaijan.\\
% ...

\end{flushleft}


%

%
%
\clearpage

\end{document}